\documentclass{an}
\usepackage{times}
\usepackage{amssymb}
\usepackage{amsmath}
\usepackage{graphicx}
\usepackage{txfonts}
\newcommand{\myr}{\,$M_{\sun}\,{\rm yr}^{-1}$}
\newcommand{\ubv}{$UBV~$}

\newcommand{\kms}{\,km\,s$^{-1}$~}
\newcommand{\ha}{H$\alpha~$}
\newcommand{\I}{\rm {\small I}}
\begin{document}
%
\Pagespan{242}{255}
\Yearpublication{2012}%
\Yearsubmission{2011}%
\Month{03}%
\Volume{333}%
\Issue{3}%
\DOI{10.1002/asna.201111655}

\title{Recent photometry of symbiotic stars -- XIII\thanks{Tables 
       2-16 are available at the CDS via
http://cdsarc.u-strasbg.fr/cgi-bin/qcat?J/AN/333/242~ 
or~ 
http://www.ta3.sk/$\sim$astrskop/symbphot/}}

\author{A.~Skopal\inst{1}\fnmsep\thanks{Corresponding author:
        \email{skopal@ta3.sk}\newline}
\and  S.~Shugarov\inst{1}
\and  M.~Va\v{n}ko\inst{1}
\and  P.~Dubovsk\'y\inst{2}
\and  S.~P.~Peneva\inst{3}
\and  E.~Semkov\inst{3}
\and  M.~Wolf\inst{4}\thanks{Visiting Astronomer, Hvar 
                             Astronomical Observatory}
}
\titlerunning{Photometry of symbiotic stars}
\authorrunning{A. Skopal et al.}
\institute{
     Astronomical Institute, Slovak Academy of Sciences,
     059\,60 Tatransk\'{a} Lomnica, Slovakia
\and
     Vihorlat Astronomical Observatory, Mierov\'a 4, 
     Humenn\'e, Slovakia 
\and
     Institute of Astronomy and National Astronomical Observatory, 
     Bulgarian Academy of Sciences, 72 Tsarigradsko Shose blvd., 
     BG-1784 Sofia, Bulgaria
\and
     Astronomical Institute, Charles University Prague, CZ-180\,00
     Praha 8, \mbox{V Hole\v{s}ovi\v{c}k\'ach} 2, The Czech Republic
}

\received{2011 Dec 11}
\accepted{2012 Feb 29}
\publonline{2012 Apr 5}

\keywords{Catalogs -- binaries: symbiotics --
          stars: individual (EG~And, Z~And, BF~Cyg, CH~Cyg, CI~Cyg, 
                             V1329~Cyg, TX~CVn, AG~Dra, Draco~C1, 
                             AG~Peg, AX~Per) --
          Techniques: photometric}

\abstract{
We present new multicolour ($UBVR_{\rm C}I_{\rm C}$) photometric 
observations of classical 
symbiotic stars, EG~And, Z~And, BF~Cyg, CH~Cyg, CI~Cyg, V1329~Cyg, 
TX~CVn, AG~Dra, Draco~C1, AG~Peg and AX~Per, carried out between 
2007.1 and 2011.9. The aim of this paper is to present new data 
of our monitoring programme, to describe the main features of 
their light curves (LC) and to point problems for their future 
investigation. The data were obtained by the method of 
the classical photoelectric and CCD photometry. 
}

\maketitle

\section{Introduction}

Symbiotic stars are interacting binary systems comprising a cool 
giant as the donor star and a compact star, mostly a white dwarf, 
as the accretor. The accretion process from the giant's wind 
heats up the accretor to $T_{\rm acc} \gtrsim 10^5$\,K and makes 
it as luminous as $L_{\rm acc} \approx 10^2 - 10^4\,L_{\sun}$. 
Such the hot and luminous source of radiation ionizes the 
circumstellar matter in the binary giving rise to a strong 
nebular emission. This basic composition of symbiotic binaries, 
containing radiative sources of extremely different temperatures 
makes the symbiotic phenomenon observable within a very large 
range of the electromagnetic spectrum, from X-rays to the radio. 
This general view have been originally pointed out by, e.g., 
Boyarchuk (1967), Allen (1984), Kenyon (1986), 
Nussbaumer \& Vogel (1987) and most recently was discussed 
during the Asiago workshop on symbiotic stars 
(Si\-vi\-ero \& Munari, 2011). 

During {\em quiescent phase}, when the symbiotic system releases 
its energy approximately at a constant rate and temperature, the 
symbiotic nebula is represented predominantly by the ionized 
fraction of the wind from the giant (e.g. Sea\-quist et al. 1984). 
A typical signature of the quiescent phase is a wave-like 
orbitally related variation in the LCs. Originally, Boyarchuk (1966), 
Belyakina (1970a) and Kenyon (1986) suggested a reflection/heating 
effect as being responsible for this type of the light variability. 
Later, Skopal (2001) interpreted the wave-like variability within 
the ionization model of symbiotic stars during quiescent phases. 
According to Skopal (2006), during {\em active phases}, the enhanced 
wind from the hot star becomes a vital source of the nebular 
radiation in the system. In addition, an optically thick warm 
($1-2 \times 10^4$\,K) source develops during the active phases 
around the hot star (e.g. Kenyon \& Webbink, 1984; Skopal, 
2005a). Location of both the radiative sources in the vicinity 
of the hot star makes them a subject to the eclipse by the giant 
in highly inclined orbits. As a result, narrow minima (eclipses) 
can be observed in the LC (e.g. Belyakina 1979). 

The observed spectrum of symbiotic stars composes of three basic 
components of radiation -- two stellar and one nebular. Their 
contributions in the optical rival each other and are different 
for different objects and variable due to activity and/or the 
orbital phase (see Figs.~2--22 of Skopal, 2005a). Therefore, 
the LCs of symbiotic binaries bear a great deal of information 
about properties of the radiative sources in the system. 
Their disentangling into the individual components of radiation 
aid us in understanding responsible physical processes acting 
in these systems (Carikov\'a \& Skopal, 2010; 
Fig.~8 of Skopal et al. 2011). 
They represents an important complement to observations carried 
out at other wavelengths, from X-rays to the radio. Their 
systematic monitoring plays an important role in discoveries 
of unpredictable outbursts of symbiotic stars, providing thus 
an alert for observation with other facilities. 

In this paper we present results of our long-term monitoring 
programme of photometric observations of selected symbiotic 
stars, originally launched by Hric \& Skopal (1989). It continues 
the work of Skopal et al. (2007, hereafter Paper~I) by collecting 
new data obtained during the period 2007 January to 2011 November. 
Their acquisition and reductions are introduced in Sect.~2. 
In Sect.~3 we describe the most interesting features of the LCs 
that deserve further investigation. 
Conclusions are found in Sect.~4. 
%
%
\begin{figure*}
\centering
\begin{center}
\resizebox{\hsize}{!}{\includegraphics[angle=-90]{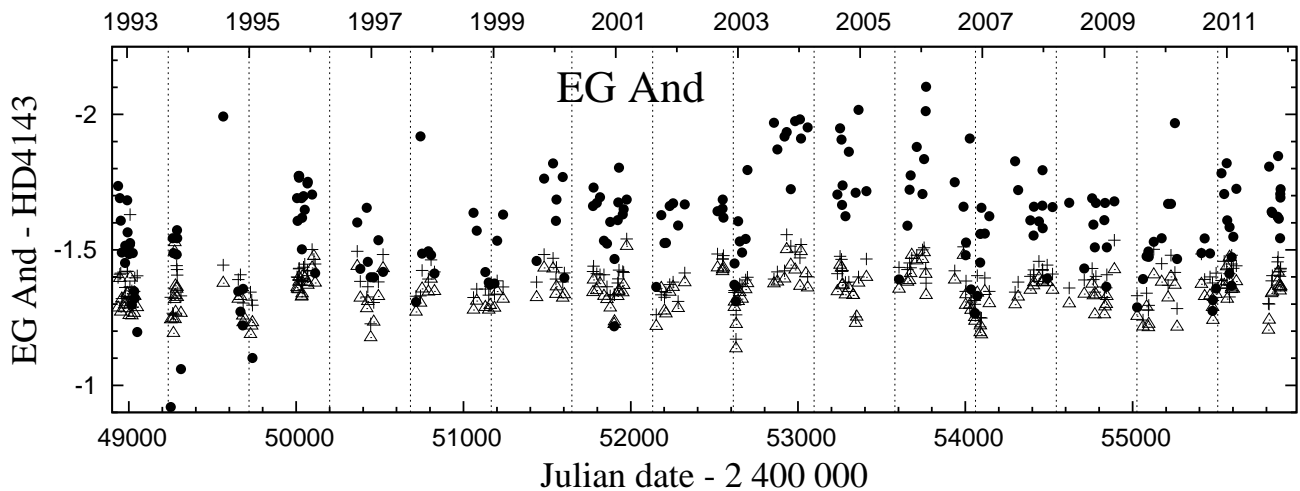}}
\caption{
$\Delta U(\bullet),\,\Delta B(\triangle),\,\Delta V(+)$ 
light curves of EG~And from 1993. Vertical dotted lines mark 
times of the inferior conjunction of the giant according to 
the ephemeris (1). 
Data are from Table~2 and previous papers of this series 
(see Paper~I and references therein). 
}
\end{center}
\end{figure*}

\section{Observations and reductions}

Photoelectric \ubv observations were carried out by 
single-channel photometers at the Skalnat\'{e} Pleso and Hvar 
observatories. The photoelectric measurements were done in 
the \ubv filters of the Johnson's photometric system with 
a 10 second integration time. 
Observations made at the Skalnat\'{e} Pleso observatory 
was described by Skopal et al. (2004). 
%

At the Hvar observatory, classical photoelectric $UBV$ 
observations were carried out by a single-channel photometer 
mounted in the Cassegrain focus of a 0.65-m reflector. 
The measurements were carefully reduced to the standard 
Johnson system via non-linear transformation formulae 
(Harmanec et al. 1994) using the latest release 17 of the 
program {\sc HEC22}~\footnote{The whole package containing 
the program {\sc HEC22} and other programs for complete 
photometric reductions, sorting and archiving the data, 
together with a very detailed User manual, is freely available 
at {\it http://astro.troja.mff.cuni.cz/ftp/hec/PHOT.} }

CCD photometry was obtained at the Star\'{a} Lesn\'{a} and 
the National Astronomical Observatory Rozhen, Bulgaria. 
At the Star\'{a} Lesn\'{a} observatory, the {\sc SBIG ST10 MXE} 
CCD camera with the chip 2184$\times$1472 pixels and the 
$UBV(RI)_C$ Johnson-Cousins filter set were mounted at the 
Newtonian focus of a 0.5-m telescope. The size of the pixel 
is 6.8\,$\mu$m and the scale 0.56$\arcsec$/pixel, corresponding 
to the field of view of a CCD frame about of 24$\times$16 arcmin. 

At the Rozhen observatory, CCD observations were made mostly with 
the 50/70/\-172\,cm Schmidt telescope. A CCD camera SBIG ST-8 and 
a Johnson-Cousins set of filters were used. The chip of the camera 
is KAF 1600 (16 bit), with dimensions of 13.8$\times$9.2\,mm or 
1530$\times$1020 pixels. The pixel size is 9$\times$9\,$\mu$m 
and the scale 1".1/pixel. 
The readout noise was 10 ADU/pixel and the gain 2.3 e-/ADU. 
All frames were dark subtracted and flat fielded. Photometry 
was made with {\scriptsize DAOPHOT} routines. 
%
Observations of V1329~Cyg and Dra\-co~C1 in the standard 
Johnson-Cou\-sins system were made with the VersArray 1300\,B 
CCD camera ($1340 \times 1300$\,px, pixel size: 
20\,$\mu$m $\times $20\,$\mu$m, scale: 0.258 arcsec/px) on 
the 2-m and 1.3-m telescopes. 

Fast CCD photometry was performed at the Star\'a Lesn\'a 
observatory (CH~Cyg) and at the Astronomical Observatory on 
the Kolonica Saddle (Z~And) with a WATEC\,902-H2/FLI\,PL1001E 
CCD camera with the chip 1024x1024 pixels attached to the 
Ritchey-Chretien telescope 300/2400 mm. The scale was 
2.06$\arcsec$/pixel corresponding to the field of view of 
a CCD frame about of 35.2$\times$32.5 arcmin. 

We measured our targets with respect to the same standard stars 
as in our previous papers (Paper~I and references therein). 
For a better availability, we summarize them here in Table~1. 
Results are summarized in Tables~2--16 and shown in Figs.~1--16. 
Each value represents the average of the observations during 
a night. This approach reduced the {\em inner} uncertainty of 
this night-means to of a few times 0.01\,mag in the $B$, $V$, 
$R_{\rm C}$ and $I_{\rm C}$ bands, and up to 0.05\,mag in 
the $U$ band. 
%
%
%
\begin{table*}
\begin{center}
\caption{Magnitudes of the comparison stars used for our targets}
\begin{tabular}{lccccccc}
\hline
\hline
Name &  V  & $B-V$ & $U-B$ & $V-R_{\rm C}$ &$V-I_{\rm C}$& Table & Refs. \\
\hline
\multicolumn{8}{c}{Comparison stars in the field of EG~And} \\
\hline
HD\,4143    & 8.574 & 1.540 &1.965 & 0.819 & 1.578 & 2,11  & 1  \\
HD\,3914$^a$& 7.00  & 0.44  & --   &  --   &  --   & 2,11  & 2  \\
HD\,4322$^a$& 7.55  & 0.47  & --   &  --   &  --   & 2,11  & 2  \\
\hline
\multicolumn{8}{c}{Comparison stars in the field of Z~And} \\
\hline
SAO\,53150    & 8.985 & 0.410 & 0.138 & 0.091 & --   & 3      & 3  \\
SAO\,53150    & 9.082 & 0.392 & 0.095 & 0.292 & 0.494&4$^b$,11& 1  \\
SAO\,53133$^a$& 9.169 & 1.360 & 1.106 &  --   & --   & 3      & 3  \\
SAO\,53133$^a$& 9.229 & 1.320 & 1.229 & 0.744 & 1.425& 11     & 1  \\
\hline
\multicolumn{8}{c}{Comparison stars in the field of BF~Cyg} \\
\hline
HD\,183650        & 6.96  & 0.71  & 0.34  &  --   & --   & 5    & 2 \\
BD+30\,3594$^a$   & 9.54  & 1.20  & 1.70  &  --   & --   & 5    & 2 \\
TYC~2137-847-1$^c$&11.159 & 0.290 & 0.091 & 0.173 & 0.381& 6    & 1 \\
\hline
\multicolumn{8}{c}{Comparison stars in the field of CH~Cyg} \\
\hline
HD\,183123    & 8.353 & 0.479 &-0.025 & 0.312 & --   & 7      & 4  \\
HD\,182691$^a$& 6.525 &-0.078 &-0.240 & 0.000 & --   & 7      & 2  \\  
BD+49\,3005   & 9.475 & 0.546 & 0.079 & 0.349 & 0.642& 8,11   & 1  \\
\hline
\multicolumn{8}{c}{Comparison stars in the field of CI~Cyg} \\
\hline
HD\,187458     & 6.660 & 0.426 &-0.056 &  --   & --   & 9      & 5  \\ 
HD\,226107$^a$ & 8.638 &-0.053 &-0.314 &  --   & --   & 9      & 6  \\ 
TYC~2861-1332-1&11.722 & 0.274 & 0.198 & 0.159 & 0.332& 10,11  & 1  \\ 
\hline
\multicolumn{8}{c}{Comparison star in the field of V1329~Cyg} \\
\hline
 "b"        &12.092 & 1.353 & 1.285 & 0.724 & 1.370& 11  & 1  \\ 
\hline
\multicolumn{8}{c}{Comparison stars in the field of TX~CVn} \\
\hline
SAO\,63223    & 9.36  & 0.30  & 0.03  &  --   & --   & 12     & 2  \\ 
SAO\,63189$^a$& 9.18  & 0.38  &-0.07  &  --   & --   & 12     & 7  \\ 
BD+37\,2314   &11.335 & 1.018 & 0.873 & 0.555 & 1.014& 11     & 8  \\
\hline
\multicolumn{8}{c}{Comparison stars in the field of AG~Dra} \\
\hline
SAO\,16952    & 9.88  & 0.56  &-0.04  &  --   & --   & 13     & 2  \\
SAO\,16935$^a$& 9.46  & 1.50  & 1.89  &  --   & --   & 13     & 7  \\
TYC~4195-369-1&10.459 & 0.559 & 0.015 & 0.333 & --   & 11,14  & 1  \\
 "b"          &11.124 & 0.734 & 0.183 & 0.416 & 0.746& 11,14  & 1  \\
 "c"          &11.699 & 0.545 &-0.042 & 0.335 & 0.629& 11,14  & 1  \\ 
\hline
\multicolumn{8}{c}{Comparison stars in the field of Draco~C1} \\
\hline
 "4"          &15.363 & 0.965 & 0.620 & 0.517 & 0.998& 11     & 9  \\
 "7"          &17.447 & 0.791 & 0.298 & 0.465 & 0.895& 11     & 9  \\
 "9"          &17.974 & 1.007 & 0.309 & 0.585 & 1.166& 11     & 9  \\  
"11"          &19.339 & 0.784 & 0.114 & 0.445 & 1.014& 11     & 9  \\
"12"          &19.787 & 0.800 &  --   & 0.407 & 1.050& 11     & 9  \\
\hline
\multicolumn{8}{c}{Comparison stars in the field of AG~Peg} \\
\hline
HD\,207933    & 8.10  & 1.05  & 0.97  &  --   & --   & 15     & 2  \\
HD\,207860$^a$& 8.73  & 0.42  & --    &  --   & --   & 15     & 2  \\
TYC~1130-577-1&10.428 & 0.631 & 0.178 & 0.374 & 0.682& 11     & 1  \\
LP~518-54     &10.672 & 0.832 & 0.528 & 0.508 & 0.938& 11     & 1  \\
\hline
\multicolumn{8}{c}{Comparison stars in the field of AX~Per} \\
\hline
BD+54\,331    & 7.427 & 1.016 & 0.632 &  --   & --   & 16     & 2  \\
BD+53\,340$^a$& 9.482 & 1.369 & 1.199 &  --   & --   & 16     & 7  \\
TYC~3671-791-1&10.201 & 0.089 &-0.177 & 0.023 & 0.066& 11     & 1  \\   
$\beta$       &11.156 & 1.167 & 0.993 & 0.608 & 1.139& 11     & 1  \\
\hline
\hline
\end{tabular}
\end{center}
Refs.: 1 -- Henden \& Munari (2006), 2 -- Blanco et al. (1970), 
      3 -- Skopal et al. (2000a), 4 -- Skopal et al. (2000b), 
      5 -- Harmanec et al. (1994), 6 -- this paper, 
      7 -- Skopal et al. (2004), 8 -- Henden \& Munari (2001), 
      9 -- Henden \& Munari (2000) \\
$^a$ -- a chech star, 
$^b$ -- for $R_{\rm C},\, I_{\rm C}$ filters, 
$^c$ -- variable star of $\delta$\,Sct type
\normalsize
\end{table*}

\section{Light curves of the measured objects}


\subsection{EG~And}

%
EG~And is a bright symbiotic binary (V$\sim 7.5)$, whose optical 
spectrum is dominated by the radiation from the cool giant 
(see Fig.~2 of Skopal, 2005a). It contains a low-luminosity
white dwarf powered by the accretion from the giant's wind 
at a rate of a few times $10^{-8}$\myr\ (Skopal, 2005b). To date, 
no outburst has been recorded. EG~And is the eclipsing binary, 
whose eclipses can be measured in the far-UV, where the hot star 
dominates the spectrum (e.g. Vogel, 1991; Pereira, 1996; Crowley 
et al. 2008). It is a near star (d = 0.59\,kpc, Skopal, 2005a) 
with a small interstellar absorption on the line of sight to it 
($E_{\rm B-V} = 0.08$, M\"{u}rset et al. 1991). It is bright in 
the far-UV, containing a strong He\I\I\,1640\,\AA\ emission line, 
when viewing the binary from its hot component (e.g. Crowley 
et al. 2008). These properties makes EG~And a good target 
for the X-ray detection around the orbital phase 0.5. 
%
%
%
\begin{figure}
\centering
\begin{center}
%
\resizebox{8cm}{!}{\includegraphics[angle=-90]{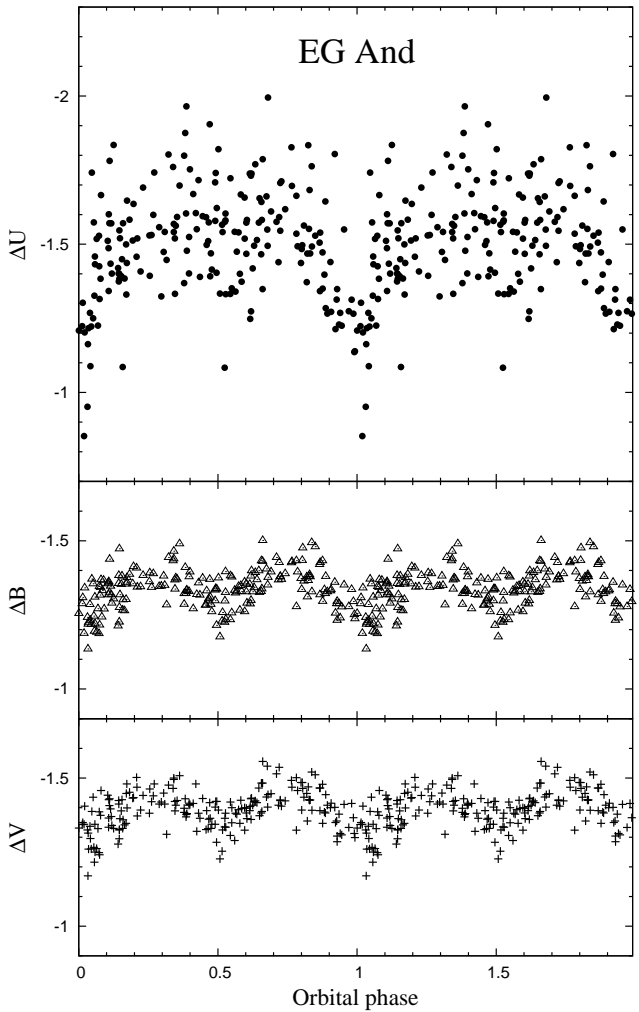}}
\caption{
Phase diagrams for the data in Fig.~1 folded with 
the ephemeris (1). 
}
\end{center}
\end{figure}
%

New photometric measurements are listed in Tables~2 and 11. 
Figure~1 shows evolution in the $UBV$ LCs during the last 18 
years, from 1993. 
Basically, two main types of the brightness variations can 
be recognized. A very slow brightening in $U$ on the time-scale 
of years with a maximum between 2003 and 2007, and shorter 
variations indicated on the time-scale of the orbital period 
(see Fig.~1). 
To demonstrate the orbitally-related variation, we folded the 
data from Fig.~1 into a phase diagram according to the ephemeris 
of the inferior conjunction of the giant (Fekel et al. 2000), 
\begin{equation}
  JD_{\rm sp.~conj.} = 2~450\,683.2(\pm 2.3) +
                  482.57(\pm 0.53) \times E. 
\end{equation}
Figure~2 shows the result. The $B$ and $V$ band LCs display 
two waves throughout the orbit with the minima around the 
giant's conjunctions, while the secondary minimum in $U$ is 
only putative due to a large scatter in the data. 
Wilson \& Vaccaro (1997) interpreted this LC profile in the $B$ 
and $V$ band by the ellipsoidal distortion of the cool giant 
filling its Roche lobe. However, 
Jurdana-\v{S}epi\'c \& Munari (2010), using 257 photographic 
measurements of EG~And between 1958.7 and 1993.0, did not find 
a pronounced minimum around the phase 0.5 in the $B$ LC 
(see their Fig.~4). Therefore they preferred an irradiation 
effect to the ellipsoidal distortion of the giant as a possible 
cause of the LC profile along the orbit. 
However, the problem is more complex, because of an intrinsic 
variability of both the giant and the nebula, which affects 
significantly the LC profile (Fig.~2). 

(i) 
Variations from the nebula are measured mainly in the $U$ band, 
where the nebular continuum dominates the spectrum. The scatter 
in the observed data is as large as $\sim$0.5\,mag, far beyond 
their uncertainties. A putative secondary minimum around the 
phase 0.5 can be caused by a partially optically thick nebula 
between the binary components as proposed by Skopal (2001) 
for this type of the variability seen in LCs of symbiotic stars 
(see also Skopal, 2008). It is of interest to compare the phase 
diagram for the $U$ magnitudes obtained before $\sim 1993$, 
where the secondary minimum in $U$ was much more pronounced 
(see Fig.1 of Skopal (1997)). 

(ii)
Intrinsic variability of the giant is best indicated in the phase 
diagrams of the $B$ and $V$ LC, where a shift of $\approx 0.1$\,mag 
is measured between the data obtained at the same orbital phases, 
but different orbital cycles (Fig.~2). 
Therefore, to understand the LC profile of EG~And, the intrinsic 
variation of the giant and the nebula has to be subtracted to 
isolate the net effect causing the double-wave variability 
throughout the orbit. 

Additional problem in explaining the LC behaviour is connected 
with finding by Jurdana-\v{S}epi\'c \& Munari (2010), whose 
archival data between 1958 and 1993 suggested a 479.28-day orbital 
period. This value is by $\sim 3.3$ days shorter than the preset 
value derived from a more recent data (Fekel et al. 2000). 
A detailed analysis in this respect with the aim to identify 
possible real change of the orbital period should be performed. 

Finally, with respect to the above mentioned results and problems, 
a future work should continue mainly the monitoring the stars's 
brightness in the $U$ band. To understand the fast variability in $U$, 
we also suggest to continue the X-ray observations (see M\"urset at 
al. 1997) around the orbital phase 0.5. 
%
%
%
%
\begin{figure*}
\centering
\begin{center}
%
\resizebox{\hsize}{!}{\includegraphics[angle=-90]{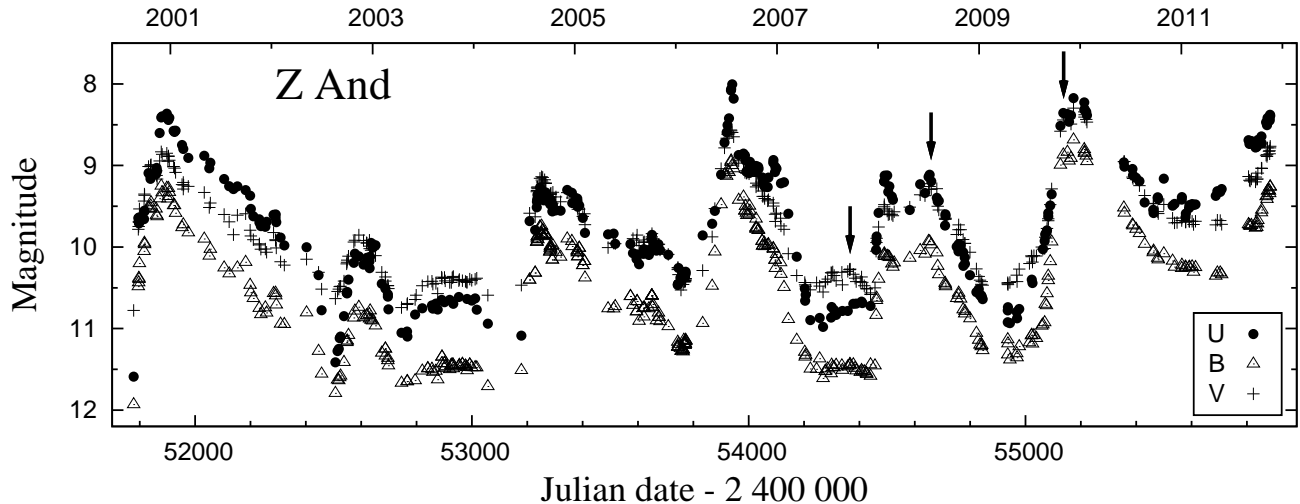}}
\caption{
The \ubv\ light curves of Z~And covering its recent active
phase from 2000.7. New data are from 2007.115. Arrows mark 
dates, when the high-time-resolution photometry was 
performed (Fig.~4). 
}
\end{center}
\end{figure*}

\subsection{Z~And} 
Z~And is considered a prototype symbiotic star. Its historical 
light curve from 1900 was recently discussed by 
Leibowitz \& Formiggini (2008). The current active phase of 
Z~And started from 2000 September (Skopal et al. 2000a), and 
continues to the present in a form of around one-year-lasting 
bursts with brightening by 1-3\,mag in $U$ (e.g. Tomov et al. 
2004; Sokoloski et al. 2006; Skopal et al. 2006; Fig.~3). 

Our new photometry of Z~And is listed in Tables~3, 4 and 11. 
Figure~3 shows \ubv\ LCs from its recent active phase from 2000. 
The star BD+47\,4192 was used as the comparison. We adopted 
its brightness and colours within the \ubv\ bands as in the 
Paper~I, while for the $R_{\rm C}$ and $I_{\rm C}$ filters we 
used colours according to Henden \& Munari (2006) 
(i.e. $V$ = 8.985, $B-V$ = 0.410, $U-B$ = 0.138, 
$V-R_{\rm C}$ = 0.195, $R_{\rm C}-I_{\rm C}$ = 0.202). 

New observations revealed two new eruptions. The first one 
appeared during 2008 with two peaks in January and July 
($U \sim$\,9.2). After the second maximum, the star's 
brightness was decreasing to 2009 May ($U \sim$\,10.9). 
The second major eruption began in 2009 August and peaked 
during 2009 December to 2010 January at $U\sim 8.2$. Then 
a fading to 2011 January ($U \sim 9.6$) and a slow gradual 
increasing to our last observations ($U\sim 8.5$ on 2011/11/18) 
was indicated. 
%
%
\begin{figure}
\centering
\begin{center}
%
\resizebox{\hsize}{!}{\includegraphics[angle=-90]{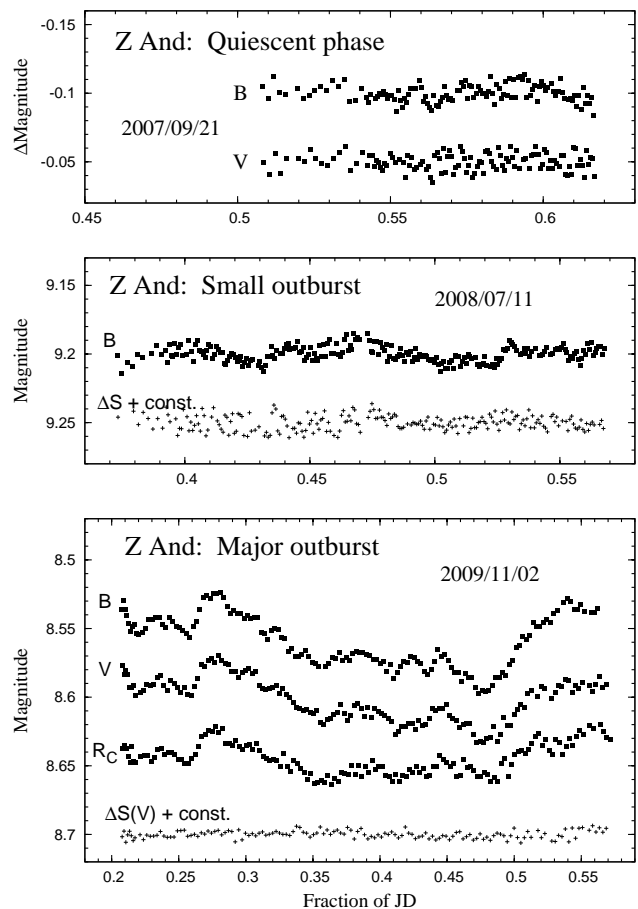}}
\caption{
Examples of high-time-resolution photometry of Z~And during 
its different levels of activity marked in Fig.~3 by arrows. 
}
\end{center}
\end{figure}
%
%
%
\begin{figure*}
\centering
\begin{center}
%
\resizebox{\hsize}{!}{\includegraphics[angle=-90]{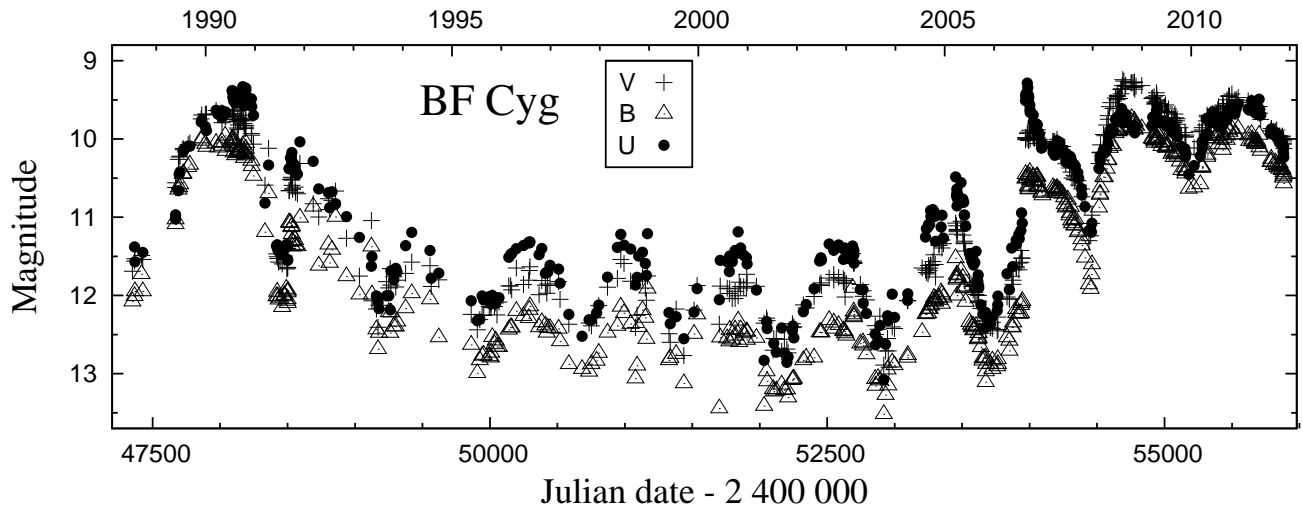}}
\caption{
The \ubv\ light curves of BF~Cyg from 1988 to the present. 
It covers the last, 1989-93 and the present, 2006-11 active 
phases with the quiescent phase in between them. 
New data are from 2007.2 (Tables~5 and 6). 
}
\end{center}
\end{figure*}

We also performed a high-time-resolution photometry (Fig.~4). 
During a low-level stage of the star's brightness, we 
indicated only a noise-like fluctuations within 0.02\,mag, 
which was comparable to that of the standard star during 
the night on 2007/09/21. 
During the smaller, 2008 burst, we observed light variations 
within an interval of $\Delta m \sim 0.025$\,mag on 
the time-scale of 1--2 hours, whereas at the maximum of the 
major 2009 outburst, the light variations increased in the 
amplitude to $\Delta m \sim 0.065$\,mag, and enlarged 
its time base to 7--9 hours, throughout the whole night. 
Here we point that such the fast photometric 
variation was also observed during the major 2006 outburst 
(see Fig.~3 of Skopal et al. 2009a), and was interpreted as 
a result of a disruption of the inner parts of the disk, 
leading to ejection of collimated mass outflow in a form of 
bipolar jets. On the other hand, during the maximum of the 
large 2000-03 outburst, no fast photometric variability was 
detected (Sokoloski et al. 2006), and no mass outflow in 
the form of jets was reported. 

Therefore, future work should be also targeted to determine 
a relationship between properties of the rapid variation and 
the star's brightness for outbursts with jets. This could aid 
us in understanding instabilities in the disk and possibly 
the mechanisms of the jet ejection.

\subsection{BF~Cyg}

The first extensive study of the LC of BF~Cyg was done 
by Jacchia (1941), by analyzing Harvard plates between 1890 and 
1940. He found a mean period of light variations of 754 days. 
Pucinskas (1970), studying variations in spectrophotometric 
parameters, derived a period of 757.3 days. 
Fekel et al. (2001) determined reliable orbital elements for 
the cool giant as 
$T_{\rm sp. conj.} = 2\,451\,395.2 \pm 5.6 + 757.2 \pm 3.9$. 
Historical light curve of BF~Cyg displays more types of 
brightening. A slow, symbiotic-nova-type outburst (1895--1960), 
outbursts of the Z~And-type (1920, 1989, 2006) and short-term 
flares (Skopal et al., 1997; Leibowitz \& Formiggini, 2006). 
The last, 1989 Z~And-type eruption was described by 
Cassatella et al. (1992) and Skopal et al. (1997). The most 
interesting feature in the LC was the development of relatively 
narrow minimum (eclipse) at the inferior spectroscopic conjunction 
of the giant, and the emergence of the P-Cyg type of profiles 
of hydrogen and helium lines. 
%
The recent, 2006 outburst of BF~Cyg was first reported by 
Munari et al. (2006). Spectroscopic observations by 
Sitko et al. (2006) and Iijima (2006a) indicated appearance 
of P-Cyg type of H\I, He\I\ and some Fe\I\I\ line profiles. 
Dramatic changes in the BF~Cyg spectrum at the beginning of 
its 2006 outburst were described by McKeever et al. (2011). 
%
%
\begin{figure}
\centering
\begin{center}
%
\resizebox{\hsize}{!}{\includegraphics[angle=-90]{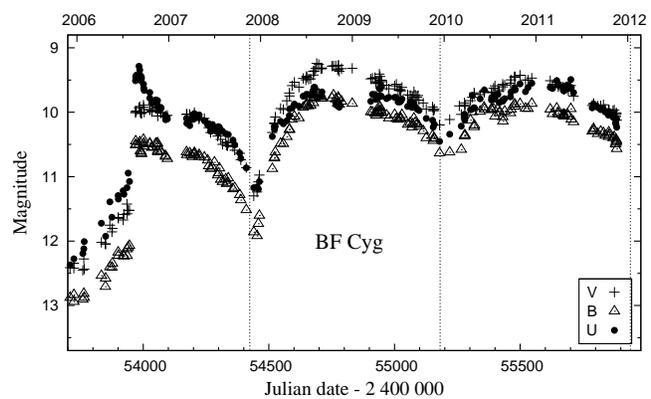}}
\caption{
A detail of the \ubv\ light curves of BF~Cyg from Fig.~5 
covering the current active phase that started in 2006 August. 
Dotted vertical lines denote times of the inferior conjunction 
of the giant according to the Fekel's et al. (2001) ephemeris. 
}
\end{center}
\end{figure}
%

The resulting night-means of the BF~Cyg brightness are given in 
Tables~5 and 6. 
For the photoelectric photometry we used the same comparison star 
as in the Paper~I. Our CCD measurements of BF~Cyg were linked to 
the comparison star "a" of Henden \& Munari (2006). 
Figure~5 shows its \ubv\ LCs covering the last 11 orbital 
cycles, from 1989. It shows that the transition from
quiescence to the present active phase began in between 
2004 and 2005 with a major eruption in 2006 August. There are 
two main differences in comparison with the previous 
1989-93 outburst. 
   (i) During the current active phase the star's brightness 
keeps ist high level continuously at/around $U\sim 10$ from the 
eruption (i.e. for $>$5 years, Fig.~6), while during the 
previous outburst, the star's brightness was fading gradually 
from its 1990 maximum at $U\sim 9.4$ to quiescent values of 
$U\gtrsim 11$ in 1993. 
   (ii) The minima (eclipses) at 2007.9 and 2010.0 varied 
in their depth and were of a V-type, in contrast to the 1991 
eclipse, which was narrower and rectangular in the profile. 
This implies changes in the geometrical structure of the hot 
component during different active phases. 
In addition, the light minima of the 2007.9 and 2010.0 eclipses 
were shifted by $\sim +20$ days with respect to the spectroscopic 
conjunction of the giant according to the Fekel's et al. (2001) 
ephemeris. 
%
%
%
\begin{figure*}
\centering
\begin{center}
\resizebox{\hsize}{!}{\includegraphics[angle=-90]{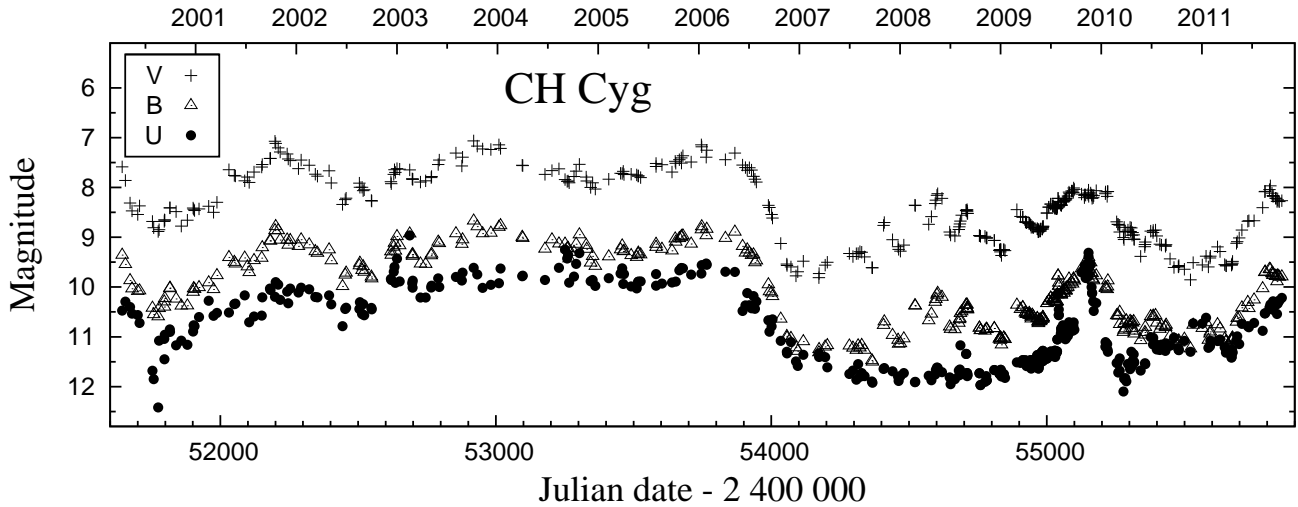}}
\caption{
 The \ubv\ light curves of CH~Cyg covering its quiescent phase 
 from 2000. 
}
\end{center}
\end{figure*}

The abnormally long-lasting high level of the BF~Cyg 
brightness requires continuation of its photometric and 
spectroscopic monitoring. The minima profiles and evolution 
in the colour indices should contribute to our understanding 
the behaviour of the hot components in symbiotic stars 
during outbursts. 
%
%
\begin{figure}
\centering
\begin{center}
\resizebox{8cm}{!}{\includegraphics[angle=-90]{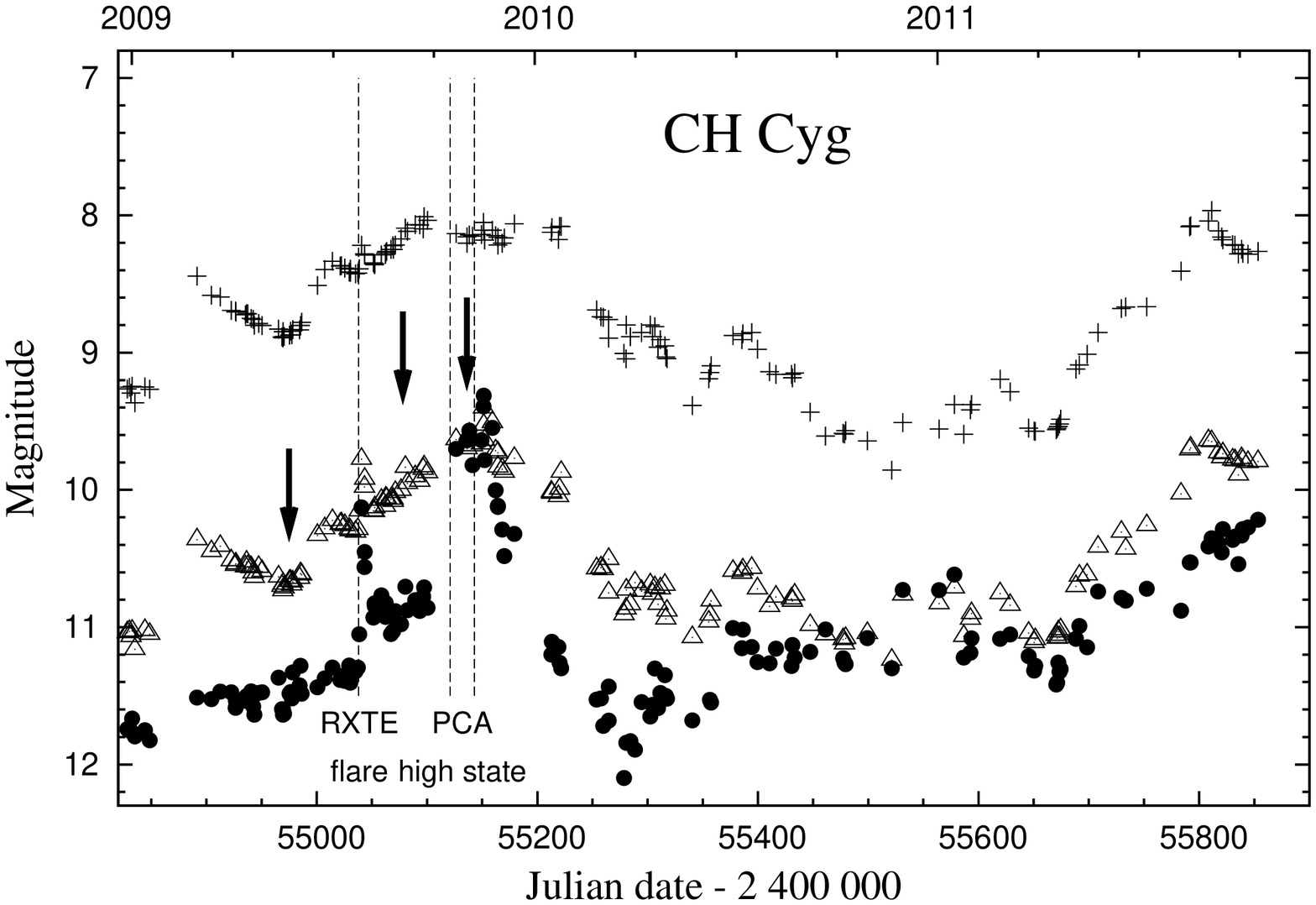}}\\[4mm]
\resizebox{8cm}{!}{\includegraphics[angle=-90]{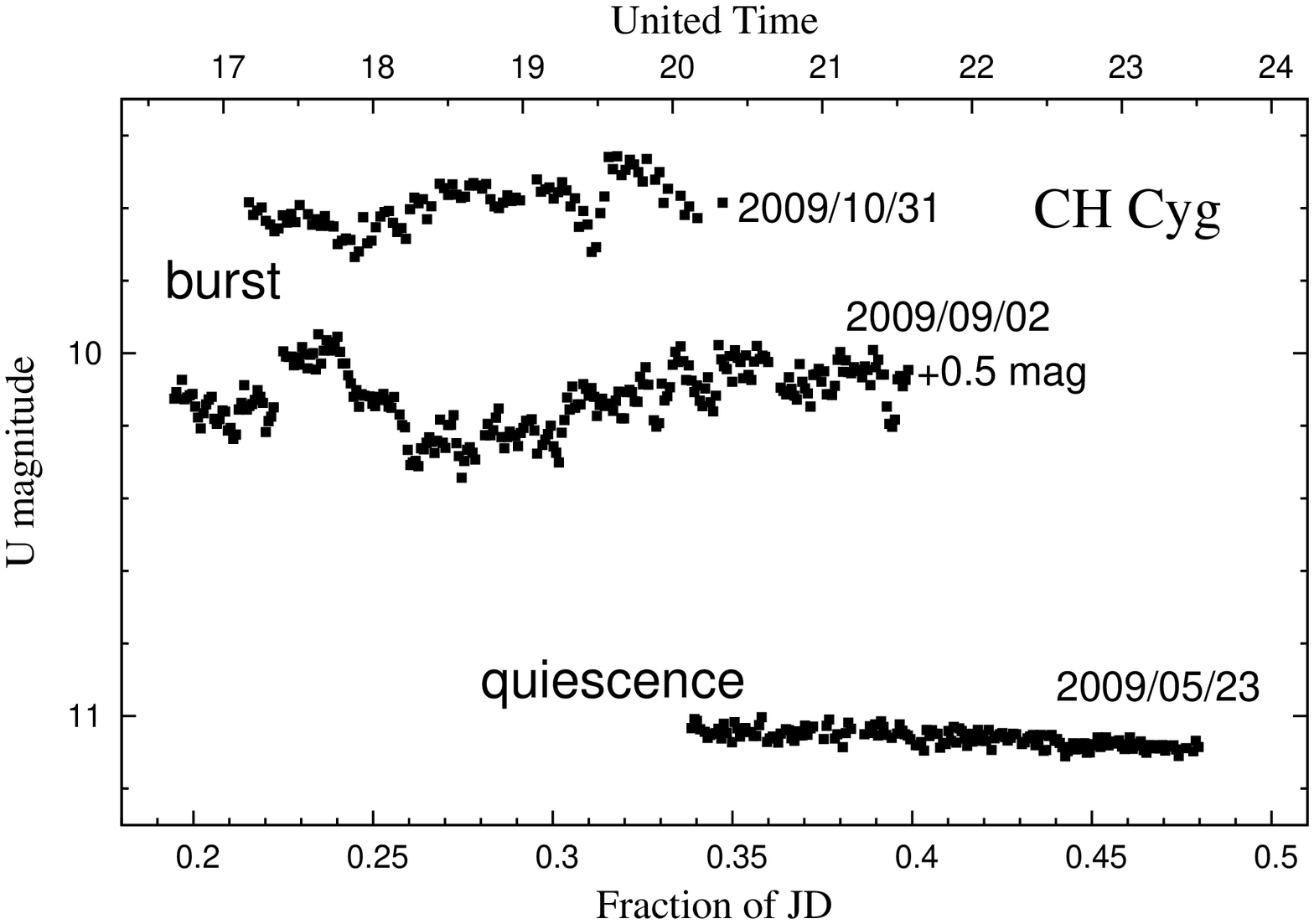}}
\caption{Top: A detail of the LCs from Fig.~7 covering sudden 
brightening in $U$ during the present low stage of CH~Cyg. 
Vertical dashed lines mark dates with a high level of the 
2--10\,keV X-ray emission. The arrows mark dates, when 
a high-time-resolution photometry was performed 
(bottom panel). 
}
\end{center}
\end{figure}

\subsection{CH~Cyg}

CH~Cyg belongs to the most intriguing symbiotic stars. Its 
symbiotic-like activity was recorded from $\sim$1963 
(Deutsch (1964); Faraggiana \& Hack (1971); 
Muciek \& Mikolajewski (1989)). Following multicolour 
observations showed that the LC-profile of CH~Cyg differs 
considerably from those of other classical symbiotic stars 
(e.g. Luud et al. (1986)). 
The object was recognized as a source of a hard X-rays 
(e.g. Leahy \& Taylor, 1987; Mukai et al. 2009a), 
producing collimated outflows in the radio 
(e.g. Taylor et al. 1986; Crocker et al., 2001; 
Karovska et al. 2007 and 2010), because of a high orbital 
inclination (e.g. Skopal et al. 1996). 
CH~Cyg is also known as a strong 
source of a fast photometric variability on a time scale from 
minutes to hours with the brightness differences of a few times 
0.1\,mag (e.g. Wallerstein, 1968; Slovak \& Africano 1978). 
Sokoloski \& Kenyon, (2003) studied changes in the rapid optical 
variability around the period of 1996-97, when jets were launched, 
and proposed that a reduction in the amplitude of the fastest 
flickering may be due to disruption of the inner disk in 
association with a mass ejection event. 
%
%
\begin{figure*}
\centering
\begin{center}
%
\resizebox{\hsize}{!}{\includegraphics[angle=-90]{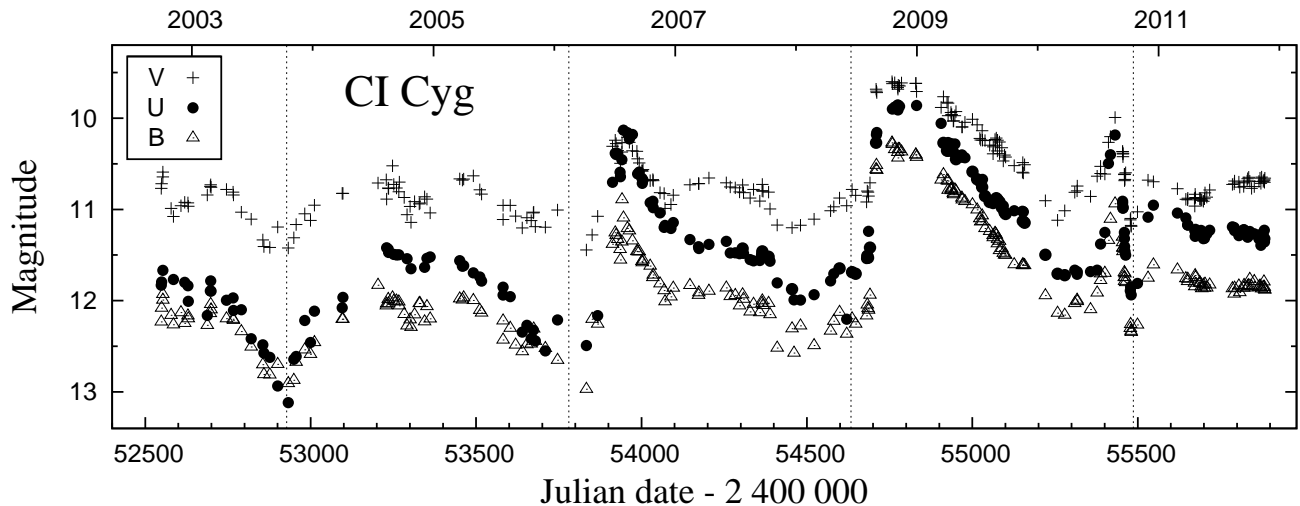}}
\caption{
The \ubv\ light curves of CI~Cyg from 2003 to the present. 
Vertical lines denote positions of the inferior conjunction 
of the giant according to the ephemeris of Fekel et al. (2000). 
}
\end{center}
\end{figure*}
%

Our new photometry of CH~Cyg is listed in Tables~7, 8 and 11. 
On the CCD snaps, CH~Cyg was measured with respect to the 
standard star "b" of Henden \& Munari (2006). Figure~7 
shows \ubv\ LCs from the beginning of 2000, which covers 
a quiescent phase with a puzzling drop to a very low optical 
state from $\sim$2006.5 to the present (Paper~I; 
Taranova \& Shenavrin, 2007). Just prior to the drop, 
emission lines in the optical spectrum were single peaked, 
but broadened at their profile bottoms to $\sim \pm 200$\kms\ 
as observed by Yoo \& Yoon (2009) on 2006 June 4. However, 
during the drop, Burmeister \& Leedj\"arv (2009) and Yoo (2010) 
observed rather strong double-peaked \ha\ profile on 2006 
October 28 and November 6, respectively. Around one year later, 
on 2007 December 29, the hydrogen emissions practically 
disappeared (Wallerstein et al. 2010). 

The photometric low state was abandoned by two short-term 
$\sim 1.5$\,mag brightening, observed during 2009 July and 
from 2009 September to the end of the year 
(Fig.~8 top; Skopal et al. 2009b and 2010). 
The burst was accompanied with appearance of a rapid light 
variability with $\Delta U \sim 0.2-0.3$\,mag, while at a low 
state ($U\sim 11.5$), the fast variations were not present 
(Fig.~8 bottom). 
Simultaneous monitoring of CH~Cyg with \textsl{RXTE} from 2007 
July (Mukai et al. 2009a and 2009b) indicated significant 
brightening in the 2--10\,keV X-ray flux during the optical 
bursts (Fig.~8 top). These independent monitoring campaigns 
revealed a correlation between the hard 2--10\,keV emission 
and that measured within the $U$ passband. 
%

Therefore, further monitoring of CH~Cyg at both the X-ray and 
the optical wavelengths is important to gain a better 
understanding of its active phases. 
%
%

\subsection{CI~Cyg}

The last active phase of CI~Cyg began in 1975 (Belyakina, 1979). 
Her photometric observations revealed deep minima in the LC, 
eclipses, which confirmed unambiguously the eclipsing nature 
of CI~Cyg. From about 1985 the profile of the minima became very 
broad, indicating transition to a quiescent phase (Belyakina, 1991). 
Multicolour $UBVRI$ observations were continued by 
Dmitrienko (1996, 2000) from 1991 to 1998. Her observations 
confirmed quiescent phase of CI~Cyg. In addition, her data, 
complemented with those of other authors, revealed a cyclic 
variability in the $U-B$ index with an amplitude of 0.3-0.4\,mag 
on a time scale of $10.7 \pm 0.6$ years. This type of the 
variability was ascribed to the hot component. 
The quiescent phase continued to 2006 May, when CI~Cyg 
started a new active phase with brightening by $ > 2$\,mag 
in $U$ (Paper~I, Fig~9 here). 
%
%
\begin{figure}
\centering
\begin{center}
%
\resizebox{8cm}{!}{\includegraphics[angle=-90]{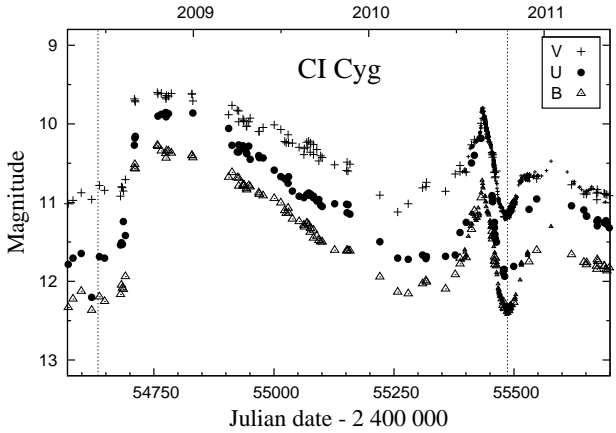}}
\caption{
The \ubv\ light curves of CI~Cyg covering its recent two 
eruptions. Vertical dotted lines are at positions of the eclipse 
timing as given by the ephemeris (2). $B$ and $V$ data were 
compared to observations from the AAVSO International Database 
(small points; Henden, 2010).
}
\end{center}
\end{figure}

Our new \ubv\ measurements of CI~Cyg are summarized in Tables~9, 
10 and 11, and are plotted in Fig.~9. 
The star HR~7550 (HD~187\,458) was used as the new comparison 
star for our photoelectric measurements. Its mean Hvar all-sky 
values, $V = 6.660, B-V = 0.426, U-B = -0.056$, were added 
to the magnitude differences, CI~Cyg--HR~7550. 
We also re-measured the previous comparison star, HD~226\,107, 
with respect to HR~7550 and obtained its new magnitudes as 
 $V = 8.638\pm 0.004$, 
 $B-V = -0.053\pm 0.008$, 
 $U-B = -0.314\pm 0.009$, 
where the uncertainties represent the rms from 23 night means 
made between 2011 May 16 and 2011 November 18. 
CCD $UBVR_{\rm C}I_{\rm C}$ magnitudes in Table~10 were obtained 
using the comparison stars "a" and "b" of Henden \& Munari (2006). 

After a gradual decrease from $U\sim 10$ in 2006 July to 
$U\sim 11.7$ in 2008 March, a drop in $U$ by $\sim 0.5$\,mag 
was observed close to the giant's inferior conjunction (Fig.~9). 
This was probably caused by a partial eclipse of the nebular 
component of radiation, because it is represented by the ionized 
wind from the hot star during active phases (Skopal, 2006). 
Following the eclipse-like effect, a strong brightening 
was measured in the mid 2008 with a maximum in $U\sim 9.8$ 
during 2008 November (Figs.~9 and 10). This eruption of CI~Cyg 
was reported by Munari et al. (2008a), and confirmed 
spectroscopically by Munari et al. (2008b) and 
Siviero et al. (2009). 
The star's brightness was again gradually decreasing to 
$U\sim 11.7$ in $\sim$2010.5, when a new fast eruption 
appeared in the LC. The 2010 bright phase was interrupted 
by the eclipse of the hot component by the giant 
(Munari et al. 2010a; Fig.~10 here). 
Our observations covered a part of the descending branch to the 
totality with the decrease rate of 0.090, 0.081 and 0.053\,mag 
a day in the $U$, $B$ and $V$ band, respectively. Our data, 
complemented with $B$ and $V$ magnitudes from the AAVSO 
database (see Fig.~10), allowed us to determine the light 
minima during the eclipse to 
  Min(B) = JD~2\,455\,487.6\,$\pm$\,1 and 
  Min(V) = JD~2\,455\.485.1\,$\pm$\,1 
by simple parabolic fit to observations between 
JD~2\,455\,457 and JD~2\,455\,515. 
We determined the middle of the eclipse as the average of 
these values, i.e., 
\begin{center}
 $JD_{\rm Ecl.}(2010.8) = 2\,455\,486.4 \pm 1.4$\,d.
\end{center}
Using the well defined timings of previous eclipses, 
\begin{center}
 $JD_{\rm Ecl.}(1975.8) = 2\,442\,691.6 \pm 0.8$, \\
 $JD_{\rm Ecl.}(1978.1) = 2\,443\,544.7 \pm 2.6$, \\
 $JD_{\rm Ecl.}(1980.4) = 2\,444\,397.9 \pm 0.8$, \\
\end{center}
(see Table~2 of Skopal, 1998) gives their ephemeris as 
\begin{equation}
  JD_{\rm Ecl.} = 2~441\,838.8(\pm 1.3) +
                  852.98(\pm 0.15) \times E.
\end{equation}
The recent observations from $\sim 2011.3$, show a more or less 
constant level of the star's brightness at $U\sim 11.3$. 
The corresponding \ubv\ LCs are similar to those observed 
after the 2006-maximum (Fig.~9). 

Further monitoring of CI~Cyg will be focused on covering the 
eclipse profile at different levels of the star's brightness 
to understand better variations of the hot component radiation 
during active phases. 
%
%
%
%
\begin{figure}
\centering
\begin{center}
%
\resizebox{\hsize}{!}{\includegraphics[angle=-90]{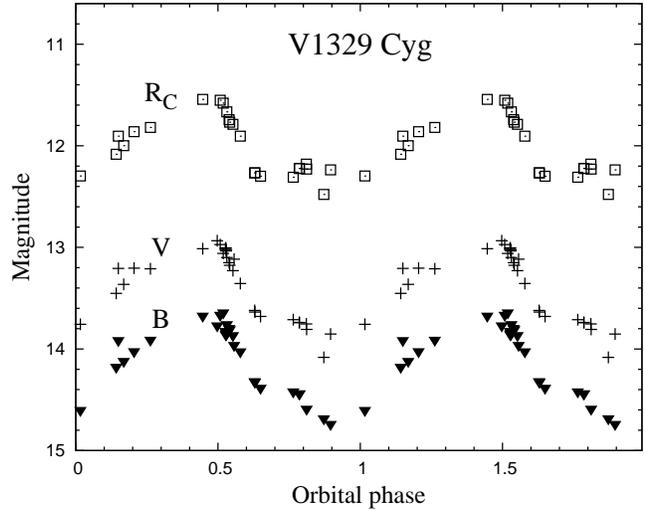}}
\caption{
Our $BVR_{\rm C}$ magnitudes of V1329~Cyg (Table~11) folded 
in the phase diagram using the ephemeris for the pre-outburst 
minima (eclipses) given by Eq.~(3). 
}
\end{center}
\end{figure}

\subsection{V1329~Cyg}

The symbiotic phenomenon of V1329~Cyg developed after its 
nova-like eruption in 1964. This was discovered by Kohoutek (1969), 
who found that its spectrum was similar to that of the well-known 
emission object, V1016~Cyg. Shortly after the eruption, from 
around 1968, V1329~Cyg peaked at $B\sim 12.0$\,mag and developed 
a wave-like orbitally-related variation in its LC (see the figure 
in Arkhipova \& Mandel, 1973), which represents a typical feature 
of the LCs of symbiotic stars during quiescent phases. This type 
of the light variability dominates the LC of V1329~Cyg to the 
present (e.g. Munari et al. 1988; Arkhipova \& Mandel, 1991; 
Skopal, 1998; Chochol et al. 1999; Jurdana-\v{S}epi\'c \& Munari 
2010; Fig~11 here). According to the model SED, it is caused by 
apparent variations in the nebular continuum as a function of 
the orbital phase (see Fig.~3 of Skopal, 2008). 

Our new observations of V1329~Cyg are given in Table~11, and 
the corresponding phased LC is plotted in Fig.~11. 
Here we used the ephemeris for the pre-outburst minima 
(eclipses) as derived by Schild \& Schmid (1997), 
\begin{equation}
 JD_{\rm Min} = 2\,427\,687(\pm 20) + 958.0(\pm 1.8)\times E .
\end{equation}
Figure~11 shows that the light minimum occurred prior to the 
time of the spectroscopic conjunction. 
As the nebular continuum dominates the near-UV/optical, the 
position and the profile of the light minimum are given by 
the shaping of the optically thick part of the nebula in 
the binary. 
Its central crossbar will be probably extended between 
the binary components, but placed asymmetrically with respect 
to the binary axis so to satisfy the minimum position 
preceding the inferior conjunction of the giant. To explain 
this effect, Skopal (1998) suggested that such the dense nebula 
can be a result of the colliding stellar winds from binary 
components as given by hydrodynamical calculations 
of Walder (1995). 
%
%
\begin{figure*}
\centering
\begin{center}
%
\resizebox{\hsize}{!}{\includegraphics[angle=-90]{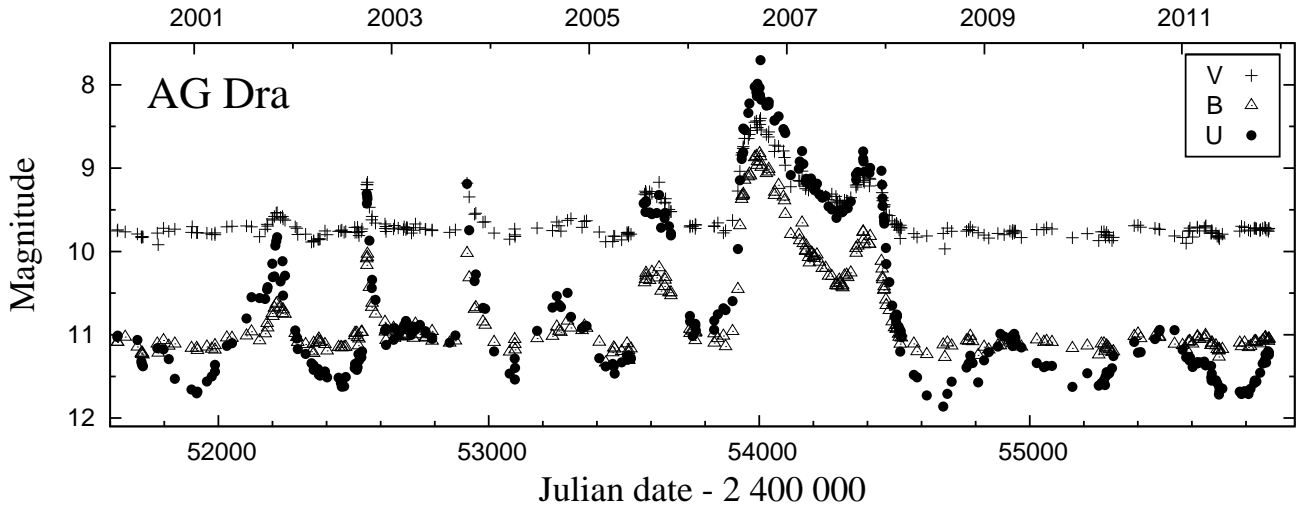}}
\caption{
The \ubv\ light curves of AG~Dra from 2000 to the present. 
 }
\end{center}
\end{figure*}
%
%
%
\begin{figure}
\centering
\begin{center}
%
\resizebox{8cm}{!}{\includegraphics[angle=-90]{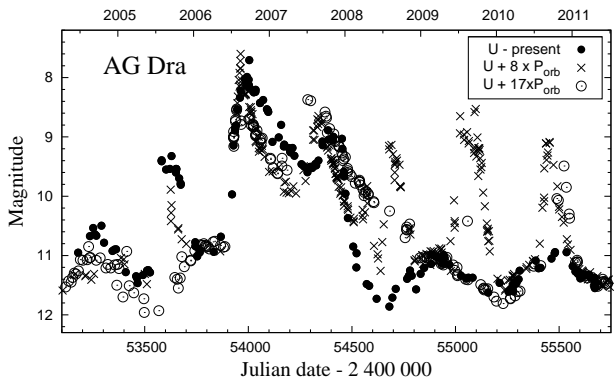}}
\caption{
Comparison of the major outbursts of AG~Dra that started 
during 1980 ($\circ$), 1994 ($\times$) and 2006 ($\bullet$). 
        }
\end{center}
\end{figure}   
%
%
%
\begin{figure}  
\centering
\begin{center}
%
\resizebox{8cm}{!}{\includegraphics[angle=-90]{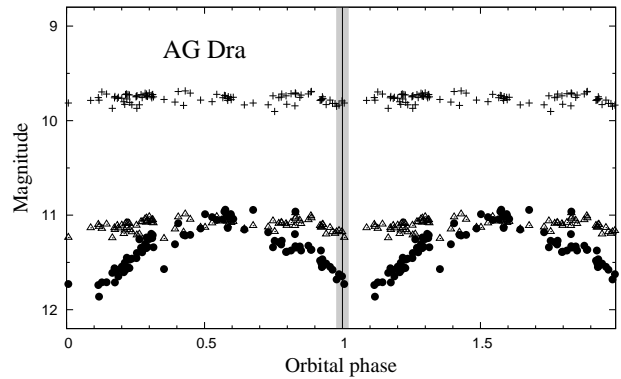}}
\caption{
Phase diagram of the $UBV$ measurements during the current 
quiescent phase from 2008 February 24 to our last observation 
on 2011 November 18 (Table~13). The width of the grey band 
represents uncertainties in the position of the inferior 
conjunction of the giant as given by elements of 
Fekel et al. (2000). 
        }
\end{center}
\end{figure}   

Figure~11 further shows that the orbital variation is well 
pronounced in all passbands ($\Delta B\sim 1.1$, 
$\Delta V\sim 0.9$ and $\Delta R_{\rm C}\sim 0.7$), whose 
amplitude is modulated by the increasing contribution from 
the red giant at longer wavelengths. 
The brightness maximum at the phase $\sim 0.5$, a rapid decrease 
to $\sim 0.65$ followed with a relatively flat minimum peaked 
at $\sim 0.9$, and finally a gradual increase to the maximum 
(see Fig.~11), could be produced by a higher density structure 
in the binary seen at different orbital phases 
(see e.g. Fig.~5 of Bisikalo et al. 2006). 

Further monitoring should be pointed to measure the precise 
profile of the LC in different colours as a function of the 
orbital motion to map the structure of the inner optically 
thick part of the nebula. Repeating X-ray observations, 
as recently performed by Stute et al. (2011), should help 
to locate the X-ray source in the binary. 
%
%

\subsection{TX~CVn}

Observations of TX~CVn are summarized in Tables~11 and 12. Since 
2008.5, monitoring of TX~CVn was finished at the Skalnat\'e Pleso 
Observatory. 
%

\subsection{AG~Dra}

The LC of the symbiotic star AG~Dra often shows multiple
brightening by 1--3\,mag, depending on the wavelength, which 
are abandoned with large periods of quiescent phases. 
First recorded activity began from 1927 with a major eruption 
($\Delta m_{\rm pg} \sim 2$\,mag) in 1951, showing a typical 
double-peaked profile (Robinson, 1964). 
First $UBV$ photoelectric observations revealed another 
major outburst around the mid of 1960' (Belyakina, 1970b). 
From 1974, AG~Dra was intensively monitored by multicolour
photometry (Meinunger, 1979; Kaler, 1987; Kaler et al. 1987).
From that time to the present, 3 major outbursts were recorded. 
They started in 1980, 1994 and 2006 (e.g. Kaler, 1980; 
Leedj\"arv et al. 2004; Paper~I). The latest one began at 
the beginning of 2006 July (Moretti et al. 2006; Iijima, 2006b). 
The spectroscopic evolution of the recent, 2006-08, outburst 
of AG~Dra was described by Munari et al. (2009a) and 
Shore et al. (2010). 
%
%
\begin{figure*}
\centering
\begin{center}
%
\resizebox{\hsize}{!}{\includegraphics[angle=-90]{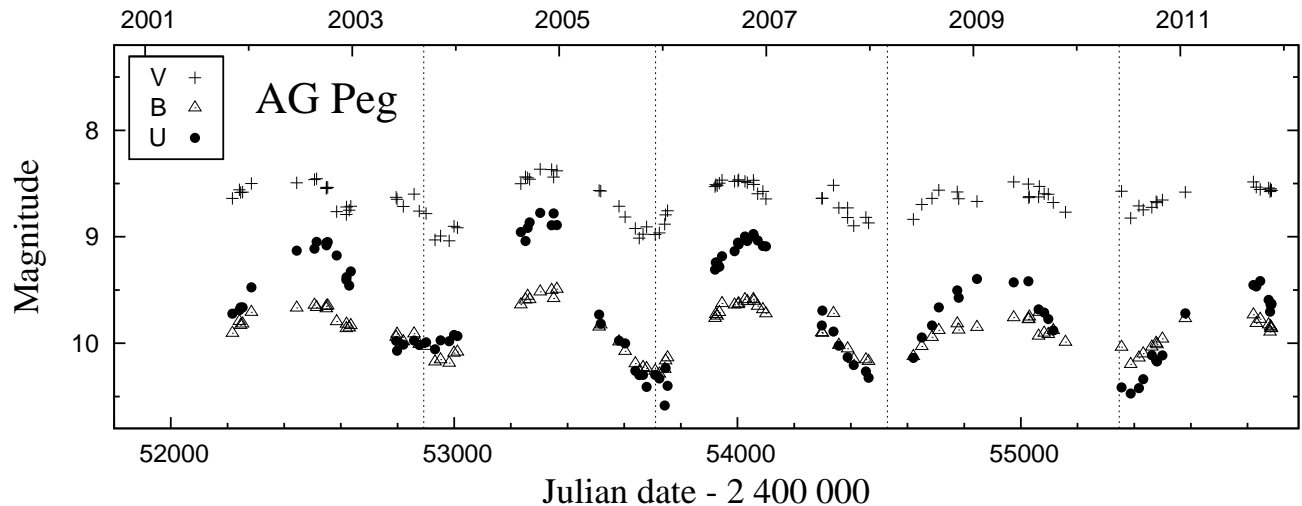}}
\caption{
The \ubv\ light curves of AG~Peg from 2002 to the present. New 
observations (Table~15) start from 2007.2. Vertical dashed 
lines denote positions of the inferior conjunction of the giant 
according to the orbital elements published by Fekel et al. (2000). 
 }
\end{center}
\end{figure*}

Our new photometry (Tables~11, 13 and 14) covers the second 
bright maximum of the 2006 outburst, which peaked during 2007 
October, being followed by a rapid decrease to quiescent 
magnitudes during 2007 December 20-th to the end of 2008 
January (Fig.~12). Transition to the quiescent phase was 
the fastest from all 3 recent outbursts. 
Another interesting feature of the AG~Dra major outbursts, 
is their beginning at nearly the same orbital phase with 
a very similar rate of the brightness increase (Fig.~13). 
The present quiescent phase was established immediately, from 
2008 February, by displaying typical wave-like orbitally-related 
variation, best pronounced in the $U$ band (Fig.~14). 
However, the position of the light minimum occurred by 
$\sim 0.05\,P_{\rm orb}$ after the position of the inferior 
giant conjunction. Maximum uncertainty in the position of 
the spectroscopic conjunction at the time of our observations 
in Fig.~14 is as large as $\sim 0.02\,P_{\rm orb}$. It is 
given by uncertainties of the Fekel et al. (2000) ephemeris, 
\begin{equation}
  JD_{\rm sp. conj.} = 2~450\,775.34(\pm 4.1) +
                  548.65(\pm 0.97) \times E, 
\end{equation}
and the average shift of 8 orbital cycles of our photometric 
observations in Fig.~14. This is in contrast to the eclipsing 
systems, where the light minima during quiescent phases precede 
the time of the inferior conjunction of the giant 
(see Skopal, 1998). 
%

\subsection{Draco~C1}

$B,~V,~R_{\rm C},~I_{\rm C}$ magnitudes of Draco~C1 
are in Table~11. We used the same standard stars as in 
the Paper~I (see Table~1). 
%
%

\subsection{AG~Peg}

AG~Peg is classified as a symbiotic nova. In mid-1850’s it rose 
in brightness from $\sim 9$ to $\sim 6$\,mag and afterwards 
followed a gradual decline to the present brightness of 8.7\,mag 
in $V$. From around 1940 the LC developed the orbitally 
related wave-like variation (Meinunger 1983). Currently, it 
displays all signatures of a classical symbiotic star in 
a quiescent phase (e.g. M\"{u}rset \& Nussbaumer, 1994). 
The spectral energy distribution in the UV/near-IR continuum 
shows a strong nebular component dominating the near-UV/U 
domain (Skopal, 2005a). 

Our new photometric observations of AG~Peg are summarized in 
Tables~11 and 15 and depicted in Fig.~15. 
The figure also compares positions of the measured light 
minima with times of the inferior conjunction of the cool 
giant in the binary, $JD_{\rm sp. conj.} = 
2\,447\,165.3(\pm 48) + 818.2(\pm 1.6)\times E$, 
as given by the elements of the spectroscopic 
orbit determined by Fekel et al. (2000). 
Variations in the minima position and their profile as well 
as the decreasing level of the star's brightness during the 
maxima suggest that the symbiotic nebula is variable in both 
the shape and the emissivity. 
For example, the $U$-LC shows a marked decrease from $\sim 9$ 
to $\sim 9.5$\,mag from the 2004.8, 2007.0 maxima to the recent, 
2009.2, 2011.5 maxima (see Fig.~15). This reflects a decrease 
in the emission measure ($EM$) of the nebular component of 
radiation by a factor of $\sim 2$. Using the method of 
Carikov\'a \& Skopal (2010), we derived 
$EM(2004.8) \sim 6.7\times 10^{59}$\,cm$^{-3}$, 
$EM(2007.0) \sim 6.6\times 10^{59}$\,cm$^{-3}$, 
$EM(2009.2) \sim 3.7\times 10^{59}$\,cm$^{-3}$ 
and 
$EM(2011.5) \sim 3.5\times 10^{59}$\,cm$^{-3}$ 
from $UBV$ magnitudes measured during the maxima. 
The magnitudes were dereddened with $E_{\rm B-V}$=0.10 
(Kenyon et al. 1993) and corrected for emission lines, 
$\Delta U_{\rm l} = 0.20$\,mag, 
$\Delta B_{\rm l} = 0.40$\,mag, 
 and 
$\Delta V_{\rm l} = 0.12$\,mag (Skopal, 2007). 
Average values of other fitting parameters, the electron 
temperature of $\approx 20\,000$\,K and the intrinsic magnitude 
of the giant, $V_{\rm g} = 8.47\pm 0.04$, were also derived. 
%
%
\begin{figure*}
\centering
\begin{center}
%
\resizebox{\hsize}{!}{\includegraphics[angle=-90]{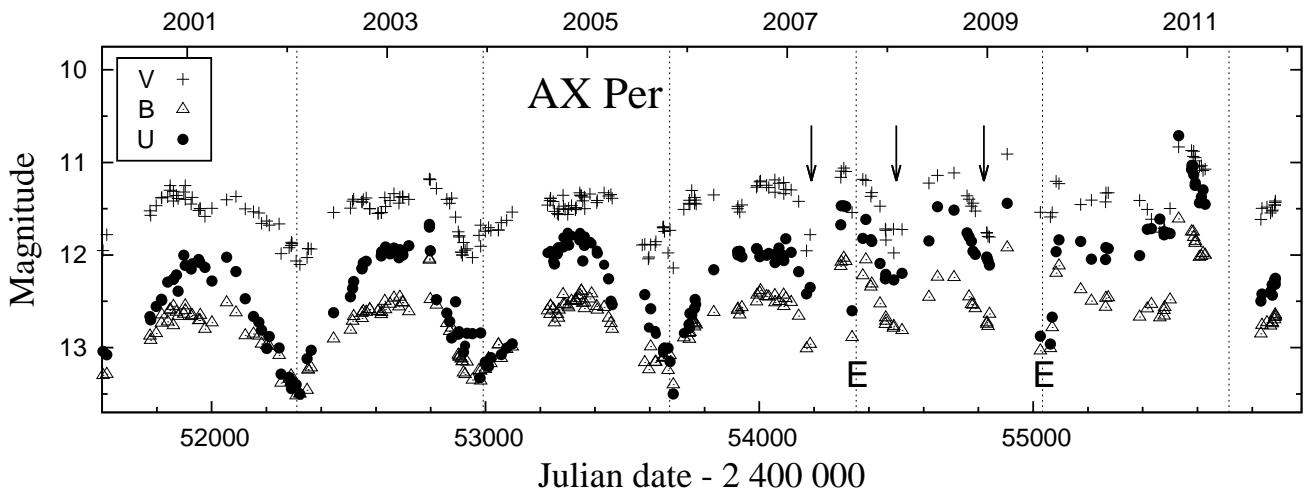}}
\caption{
The \ubv\ light curves of AX~Per from 2000 to the present.  
New observations (Table~16) are from 2007.2. Vertical dotted 
lines denote positions of eclipses according to their ephemeris 
(5) and arrows denote the minima of a wave that appeared 
in the LC during the 2007-10 active phase. 
 }
\end{center} 
\end{figure*}

Further investigation of AG~Peg should include also its soft 
and super-soft X-ray domain as previously performed by 
M\"urset et al. (1997). The small value of the interstellar 
absorption to AG~Peg and a very high level of its far-UV continuum, 
as measured by the FUSE and HST satellites, makes AG~Peg 
a suitable target for detection of the super-soft X-rays. 

\subsection{AX~Per}

AX~Per is an eclipsing symbiotic binary, whose eclipses in 
the multicolour LC were recorded for the first time during 
its major 1988-95 active phase (Skopal, 1991). 
Transition to quiescence happened during 1995, when the LC 
profile changed to the wave-like orbitally related variation 
(Skopal et al. 2001). The binary comprises a M4.5\,III giant 
(M\"urset \& Schmid 1999) and a WD on a 680-d orbit 
(e.g. Fekel et al. 2000). 

The recent measurements of AX~Per are introduced in Tables~11 
and 16, and are displayed in Fig.~16. The main change in 
the LC profile, covered by our new observations, happened at 
$\sim$2007.5, when the broad wave-like minimum transformed 
into a narrow minimum placed at the inferior conjunction of 
the giant (Fig.~16 here; Skopal et al. 2011). 
This sudden and significant change in the minimum profile 
reflects a strong change in the ionization structure of the 
binary as a consequence of a new active phase. It is interesting 
to note that this change was connected with only a small, 
a few times 0.1\,mag, increase in the optical brightness 
(see Fig.~16). 
Later on during 2009 March, a rapid increase in the star's 
brightness by $\sim$1\,mag in $B$ was reported by Munari et al. 
(2009b). They also pointed out that the observed increase of 
the nebular continuum, indicated in their spectra, was 
responsible for the increase in the star's brightness 
and the bluer colour indices. This finding confirmed that 
AX~Per was inhering in the active phase. Unfortunately, 
our scanty \ubv\ observations during that time (near to the 
season gap) did not allowed us to measure the corresponding 
brightness in $U$. The brightening was interrupted by the eclipse 
in the middle of 2009, when $U$ decreased to $\sim 13$, and 
continued at $U \sim 11.7$ after it (see Fig.~16). An additional 
rise in the AX~Per brightness during 2010 November was reported 
on by Munari et al. (2010b). Our observations confirmed the 
brightening by indicating a maximum at $U = 10.7$ on 2010 
November 30-th and the following gradual fading to 
$U \sim 11.45$ on 2011 March 3. The last observations from 
2011 September -- November, indicates a gradual increase 
from $U\sim 12.5$ to $\sim 12.2$. 

In addition, a pronounced light wave with a period of 
$\sim 0.5\,P_{\rm orb}$, broad minima located around the orbital 
phases 0.2 and 0.7 and amplitude of $\sim 0.6-0.8$\,mag, 
modulated the LC during the 2007-10 active phase (Fig.~16 here 
and Fig.~2 of Skopal et al. 2011). 
It was seen in all filters. This feature was observed in 
a symbiotic system for the first time. 
Understanding the nature of such type of variability requires 
a detailed analysis of colour indices and spectroscopic 
observations, which is out of the scope of this paper. 

Finally, the profile of the 2009 eclipse, closely covered by 
multicolour measurements obtained within the Asiago Novae and 
Symbiotic Stars Collaboration, allowed Skopal et al. (2011) 
to determine the ephemeris for eclipses in AX~Per as, 
\begin{equation}
  JD_{\rm Ecl.} = 2~447\,551.26(\pm 0.3) +
                  680.83(\pm 0.11) \times E.
\end{equation}

Further photometric and spectroscopic monitoring is important 
to map the current active phase, whose evolution of the LC is 
unusual in comparison with other active symbiotic stars. 

\section{Conclusions}

In this paper we presented new multicolour photometry of selected 
classical symbiotic stars (Tables~2--16, Figures~1--16). Main 
results of our monitoring programme can be summarized as follows. 

\ubv\ LCs of {\bf EG~And} display a double-wave throughout the 
orbit with the minima around the giant's conjunctions. The 
variation are strongly affected by an intrinsic variability 
from the giant and the nebula. For future work we suggest to 
continue monitoring the star mainly in the $U$ band, and to 
carry out X-ray observations covering the super-soft part of 
the spectrum (Sect.~3.1). 

The LC of {\bf Z~And} showed two new eruptions that peaked at 
$U\sim 9.2$ during 2008 January and July, and at $U\sim 8.2$ 
in 2009 December. During 2011 Z~And was again gradually    
brightening up. Our last observations indicated $U\sim 8.5$. 
The short-term variations with $\Delta m \sim 0.025$\,mag on 
the time-scale of 1--2 hours during the smaller outburst changed 
to $\Delta m \sim 0.065$\,mag through a night during the major 
2009 outburst (Fig.~4). 
Future work should include the high-time-resolution 
photometry with the aim to determine a relationship between 
the properties of the rapid variations and the star's 
brightness (Sect.~3.2). 

{\bf BF~Cyg} keeps a high level of its brightness at $U\sim 10$ 
from the main eruption in 2006 August. The LC was wave-like 
in the profile with minima (eclipses) of different depth 
in 2007.9 and 2010.0. Owing to the nearly rectangular profile 
of the 1991 eclipse, current evolution in the LC implies 
changes in the geometrical structure of the hot component 
during different active phases. Therefore, continuation of 
the photometric and spectroscopic monitoring of BF~Cyg is 
important to understand better the behaviour of the hot 
components in symbiotic stars during outbursts. 

{\bf CH~Cyg} persisted at a low level of its brightness with 
$U\sim 10.8 - 11.8$. The low state was abandoned with two 
short-term, $\Delta U \sim 1.5$\,mag bursts, measured in 2009 
July and 2009 September -- December. The $U$-LC correlates with 
that measured in the 2--10\,keV X-ray fluxes. 
Further monitoring of CH~Cyg at both the X-ray and the optical 
wavelengths is important to gain a better understanding of its 
active phases.

{\bf CI~Cyg} continued its active phase from 2006. A strong 
brightening was measured in the middle of 2008 with a maximum 
in $U\sim 9.8$ during 2008 November and additional smaller 
eruption that peaked at $U\sim 10.0$ in 2010 September. After 
12 orbital cycles ($\sim 28$ years), a new eclipse of the hot 
component by its giant companion, appeared in the LC during 2010 
October. We determined a new ephemeris of eclipses (Eq.~(2)). 
CI~Cyg continues its active phase at $U\sim 11.3$. 

The phased $BVR_{\rm C}$ LCs of {\bf V1329~Cyg} showed a well 
pronounced orbital variation with a complicated profile. 
Monitoring the precise profile of the LC along the orbit and 
its variation during the following orbital cycles will map 
the structure of the inner optically thick part of the nebula 
and its evolution. 

Our new photometry of {\bf AG~Dra} covered the second bright maximum 
of the 2006 outburst, which peaked during 2007 October. A rapid 
decrease happened between 2007 December 20-th and the end of 2008 
January, after which a quiescent phase was established. From 2008 
February to 2011 November, the LC displayed the wave-like 
orbitally-related variation -- a signature for quiescent 
phase of symbiotic stars. 

{\bf AG~Peg} LC continues the wave-like variation as a function of 
the orbital phase. The maximum brightness at $U\sim 9.0$, 
measured during 2004.8 and 2007.0, decreased to $U\sim 9.5$ 
during 2009.2 and 2011.5 maxima. This change was caused 
by a decrease in the emission measure of the nebula, from 
$EM \sim 6.7$ to $\sim 3.5\times 10^{59}$\,cm$^{-3}$. 

{\bf AX~Per} entered a new active phase from $\sim 2007.5$, when 
narrow eclipses around 2007.7 and 2009.5 were observed in 
the LC. However, the star's brightness increased by only 
a few times 0.1\,mag. In addition, a pronounced light wave 
with a period of $\sim 0.5\,P_{\rm orb}$ and minima located 
around the orbital phases 0.2 and 0.7 modulated the LC. 
This anomalous type of active phase requires urgently further 
photometric and spectroscopic monitoring. 
\acknowledgements
The authors thank to Ing. Jozef Lopatovsk\'y, Mr. Peter Gerberi, 
Dr. Richard Kom\v{z}\'{\i}k, Dr. Jozef \v{Z}i\v{z}\v{n}ovsk\'y, 
Dr. Lubom\'{\i}r Hamb\'alek and Dr. Matej Seker\'a\v{s} for 
acquisition of photometric observations at the Skalnat\'{e} Pleso 
observatory. 
We acknowledge with thanks the variable star observations from
the AAVSO International Database contributed by observers 
worldwide and used in this research. 
The research of M.W. was also supported by the Research Program 
MSM0021620860 of the Ministry of Education of the Czech Republic. 
This research was also supported by a grant of the Slovak Academy 
of Sciences, VEGA No. 2/0038/10. 
%
%
%

%
%
%
%
%
\begin{table*}
\begin{center}
\caption{Photoelectric~~$U,B,V,R_{\rm C}$ observations of EG~And 
         ($\Delta m$ = EG~And -- HD4143) from the Skalnat\'e Pleso
         observatory}
\begin{tabular}{lccccc}
\hline
\hline
  Date & JD~24... &$\Delta U$ & $\Delta B$ & $\Delta V$ &$\Delta R_{\rm C}$\\
\hline
 Feb 11, 2007 & 54143.254 & -1.624 & -1.302 & -1.347 & -1.448 \\
 Jul 16, 2007 & 54297.510 & -1.827 & -1.298 & -1.344 & -1.510 \\
 Aug  3, 2007 & 54315.565 & -1.721 & -1.326 & -1.381 & -1.503 \\
 Oct 15, 2007 & 54388.619 & -1.609 & -1.351 & -1.402 & -1.480 \\
 Nov  2, 2007 & 54407.480 & -1.553 & -1.368 & -1.408 & -1.534 \\
 Nov  5, 2007 & 54410.428 & -1.659 & -1.379 & -1.426 & -1.546 \\
 Dec  5, 2007 & 54440.443 & -1.605 & -1.395 & -1.458 & -1.566 \\
\hline
\hline
\end{tabular}
\end{center}
\normalsize
\end{table*}
%
%
%
\begin{table*}
\begin{center}
\caption{Photoelectric~~$U,B,V,R_{\rm C}$ observations of Z~And 
         from the Skalnat\'e Pleso observatory.}
\begin{tabular}{lccccc}
\hline
\hline
Date & JD~24... & $U$ &  $B$   &  $V$ &$\Delta R_{\rm C}$ \\
\hline
 Mar 14, 2007 & 54173.597 & 10.118 & 11.141 & 10.349 &  0.158 \\
 Apr 12, 2007 & 54202.586 & 10.510 & 11.288 & 10.434 &  0.153 \\
 Apr 13, 2007 & 54203.579 & 10.660 & 11.348 & 10.463 &  0.219 \\
 Apr 15, 2007 & 54205.585 & 10.587 & 11.327 & 10.482 &  0.231 \\
 May  2, 2007 & 54222.537 & 10.895 & 11.495 & 10.525 &  0.320 \\
 Jun  5, 2007 & 54256.523 & 10.870 & 11.371 & 10.480 &  0.325 \\
 Jun 17, 2007 & 54268.514 & 10.980 & 11.612 & 10.551 &  0.433 \\
\hline
\end{tabular}
\end{center}
\end{table*}
\addtocounter{table}{-1}
\begin{table*}
\begin{center}
\caption{Continued}
\begin{tabular}{lccccc}
\hline
Date        & JD~24...   &    $U$ &   $B$  & $V$ &$\Delta R_{\rm C}$\\
\hline
 Sep  1, 2009 & 55075.510 &  9.820 & 10.719 &  9.890 & --     \\
 Sep  7, 2009 & 55082.470 &  9.583 & 10.357 &  9.654 & --     \\
 Sep 19, 2009 & 55094.463 &  9.352 &  9.937 &  9.288 & --     \\
 Oct 31, 2009 & 55136.266 &  8.356 &  8.870 &  8.441 & --     \\
 Nov 22, 2009 & 55157.520 &  8.463 &  8.945 &  8.418 & --     \\
 Nov 28, 2009 & 55164.353 &  8.385 &  8.911 &  8.492 & --     \\
 Dec  7, 2009 & 55173.403 &  8.174 &  8.686 &  8.297 & --     \\
\hline
\hline
\end{tabular}
\end{center}
\normalsize
\end{table*}
\clearpage
%
%
\begin{table*}
\begin{center}
\caption{CCD~~$U,B,V,R_{\rm C},I_{\rm C}$ observations of Z~And 
         from the Star\'a Lesn\'a observatory.}
\begin{tabular}{lcccccc}
\hline
\hline
   Date      & JD~24... & $U$ &  $B$   &  $V$ & $R_{\rm C}$ & $I_{\rm C}$ \\
\hline
 May 25, 2007 & 54245.506 &  --    & 11.500 & 10.438 &  9.049 &  7.723 \\
 Jun 25, 2007 & 54276.510 &  --    & 11.526 & 10.432 & --     &   --   \\
 Jun 26, 2007 & 54277.514 &  --    & 11.521 & 10.425 & --     &   --   \\
 Jul  3, 2007 & 54285.366 &  --    & 11.493 & 10.393 &  9.042 &  7.744 \\
 Jul 22, 2007 & 54303.578 &  --    & 11.475 & 10.306 &  9.008 &  7.711 \\
 Aug 23, 2007 & 54336.390 &  --    & 11.456 & 10.309 &  9.004 &  7.750 \\
 Sep 21, 2007 & 54364.521 &  --    & 11.455 & 10.278 & --     &   --   \\
\hline
\hline
\end{tabular}
\end{center}
\normalsize
\end{table*}
\clearpage
%
%
%
\begin{table*}
\begin{center}
\caption{Photoelectric~~$U,B,V,R_{\rm C}$ observations of BF~Cyg 
         from the Skalnat\'e Pleso observatory.}
\begin{tabular}{lccccc}
\hline
\hline
 Date & JD~24... & $U$ &  $B$   &  $V$ &$\Delta R_{\rm C}$ \\
\hline
 Mar 13, 2007 & 54172.593 & 10.098 & 10.619 & 10.033 &  2.788 \\
 Mar 14, 2007 & 54173.526 & 10.172 & 10.663 & 10.028 &  2.792 \\
 Apr  5, 2007 & 54195.542 & 10.087 & 10.693 & 10.091 &  2.807 \\
 Apr 12, 2007 & 54202.544 & 10.100 & 10.665 & 10.085 &  2.792 \\
 Apr 12, 2007 & 54203.499 & 10.096 & 10.644 & 10.048 &  2.775 \\
 May  1, 2007 & 54222.450 & 10.086 & 10.691 & 10.107 &  2.831 \\
 Jun  4, 2007 & 54256.449 & 10.194 & 10.747 & 10.172 &  2.889 \\
\hline
\end{tabular}
\end{center}
\end{table*}
\addtocounter{table}{-1}
\begin{table*}
\begin{center}
\caption{Continued}
\begin{tabular}{lccccc}
\hline
Date        & JD~24...   &    $U$ &   $B$  & $V$ &$\Delta R_{\rm C}$\\
\hline
 Mar 25, 2010 & 55280.588 & 10.062 & 10.361 &  9.884 & --     \\
 Apr 16, 2010 & 55302.551 &  9.960 & 10.125 &  9.820 & --     \\
 Apr 23, 2010 & 55309.527 &  9.740 &  9.997 &  9.768 & --     \\
 Apr 30, 2010 & 55316.528 &  9.930 & 10.230 &  9.698 & --     \\
 May  1, 2010 & 55317.500 &  9.937 & 10.203 &  9.717 & --     \\
 May 24, 2010 & 55340.518 &  9.792 &  9.989 &  9.726 & --     \\
 Jun  8, 2010 & 55356.381 &  0.000 &  9.925 &  9.676 & --     \\
\hline
\hline
\end{tabular}
\end{center}
\normalsize
\end{table*}
\clearpage
%
%
%
\begin{table*}
\begin{center}
\caption{CCD~~$U,B,V,R_{\rm C},I_{\rm C}$ observations of BF~Cyg 
         from the Star\'a Lesn\'a observatory.}
\begin{tabular}{lcccccc}
\hline
\hline
   Date      & JD~24... & $U$ &  $B$   &  $V$ & $R_{\rm C}$ & $I_{\rm C}$ \\
\hline
 Aug 10, 2004 & 53228.385 & 11.253 & 12.242 & 11.675 & 10.370 &  9.227 \\
 Aug 29, 2004 & 53247.478 & 11.126 & 12.230 & 11.724 & 10.401 &  9.463 \\
 Sep  6, 2004 & 53255.400 &   --   & 12.173 & 11.683 & 10.370 &  9.393 \\
 Sep 19, 2004 & 53268.330 & 10.990 & 12.128 & 11.613 & 10.429 &  9.357 \\
 Oct  3, 2004 & 53282.397 & 10.903 & 12.073 & 11.567 & 10.341 &  9.537 \\
 Oct 21, 2004 & 53300.322 & 10.987 & 12.083 & 11.604 & 10.431 &  9.519 \\
 Mar 31, 2005 & 53460.627 & 10.690 & 11.737 & 11.184 & 10.039 &  9.173 \\
\hline
\end{tabular}
\end{center}
\end{table*}
\addtocounter{table}{-1}
\begin{table*}
\begin{center}
\caption{Continued}
\begin{tabular}{lcccccc}
\hline
  Date     & JD~24...   &   $U$ &   $B$  & $V$ &$R_{\rm C}$ &$I_{\rm C}$\\
\hline
 Jul 29, 2008 & 54677.356 &  9.661 &  9.856 &  9.422 &  8.975 &  8.435 \\
 Aug  3, 2008 & 54682.430 &  9.613 &  9.771 &  9.359 &  9.017 &  8.439 \\
 Aug 11, 2008 & 54690.300 &   --   &  9.696 &  9.241 &  8.952 &  8.293 \\
 Aug 19, 2008 & 54698.494 &   --   &  9.722 &  9.259 &  8.956 &  8.353 \\
 Aug 29, 2008 & 54707.524 &   --   &  9.761 &  9.325 &  8.908 &  8.403 \\
 Oct 11, 2008 & 54751.233 &   --   &  9.782 &  9.284 &  8.970 &  8.304 \\
 Oct 17, 2008 & 54757.192 &   --   &  9.735 &  9.281 &  8.906 &  8.218 \\
\hline
\hline
\end{tabular}
\end{center}
\normalsize
\end{table*}
\clearpage
%
%
%
\begin{table*}
\begin{center}
\caption{Photoelectric~~$U,B,V,R_{\rm C}$ observations of CH~Cyg 
         from the Skalnat\'e Pleso observatory.}
\begin{tabular}{lccccc}
\hline
\hline
 Date & JD~24... & $U$ &  $B$   &  $V$ &$\Delta R_{\rm C}$ \\
\hline
 Mar 12, 2007 & 54172.496 & 11.329 & 11.319 &  9.813 & -0.278 \\
 Mar 13, 2007 & 54173.482 & 11.404 & 11.254 &  9.729 & -0.440 \\
 Apr  4, 2007 & 54195.470 & 11.408 & 11.202 &  9.558 & -0.620 \\
 Apr 12, 2007 & 54203.369 & 11.614 & 11.173 &  9.503 & -0.686 \\
 Jul  1, 2007 & 54283.494 & 11.747 & 11.180 &  9.327 & -0.832 \\
 Jul 14, 2007 & 54296.438 & 11.678 & 11.223 &  9.386 & -0.801 \\
 Jul 26, 2007 & 54308.383 & 11.861 & 11.268 &  9.329 & --     \\
\hline
\end{tabular}
\end{center}
\end{table*}
\addtocounter{table}{-1}
\begin{table*}
\begin{center}
\caption{Continued}
\begin{tabular}{lccccc}
\hline
  Date     & JD~24...  &  $U$ &   $B$  & $V$ &$\Delta R_{\rm C}$ \\
\hline
 Nov 28, 2009 & 55164.299 & 10.116 &  9.833 &  8.217 & --     \\
 Dec 13, 2009 & 55179.248 & 10.321 &  9.770 &  8.062 & --     \\
 Jan 16, 2010 & 55212.659 & 11.199 & 10.014 &  8.123 & --     \\
 Mar  9, 2010 & 55264.623 & 11.681 & 10.751 &  8.894 & --     \\
 Mar 22, 2010 & 55278.496 & 12.099 & 10.909 &  9.006 & --     \\
 Mar 24, 2010 & 55280.456 &  0.000 & 10.866 &  9.045 & --     \\
 Mar 28, 2010 & 55284.489 & 11.830 & 10.835 &  8.884 & --     \\
\hline
\end{tabular}
\end{center}
\end{table*}
\addtocounter{table}{-1}
\begin{table*}
\begin{center}
\caption{Continued}
\begin{tabular}{lccccc}
\hline
  Date     & JD~24...  &  $U$ &   $B$  & $V$ &$\Delta R_{\rm C}$ \\
\hline
 Aug 10, 2011 & 55783.524 & 10.880 & 10.030 &  8.406 & --     \\
 Aug 17, 2011 & 55791.426 & 10.529 &  9.708 &  8.084 & --     \\
 Aug 18, 2011 & 55792.390 & 10.528 &  9.695 &  8.077 & --     \\
 Sep  3, 2011 & 55808.481 & 10.413 &  9.641 &  8.042 & --     \\
 Sep  6, 2011 & 55811.312 & 10.352 &  9.649 &  7.966 & --     \\
 Sep 12, 2011 & 55817.488 & 10.389 &  9.729 &  8.113 & --     \\
 Sep 15, 2011 & 55820.429 & 10.456 &  9.765 &  8.159 & --     \\
 Sep 16, 2011 & 55821.485 & 10.286 &  9.736 &  8.179 & --     \\
 Sep 25, 2011 & 55830.422 & 10.364 &  9.782 &  8.213 & --     \\
 Sep 27, 2011 & 55832.388 & 10.341 &  9.787 &  8.219 & --     \\
 Sep 30, 2011 & 55835.480 & 10.541 &  9.891 &  8.281 & --     \\
 Oct  3, 2011 & 55838.439 & 10.333 &  9.771 &  8.248 & --     \\
 Oct  4, 2011 & 55839.440 & 10.285 &  9.786 &  8.276 & --     \\
 Oct  9, 2011 & 55844.360 & 10.274 &  9.799 &  8.283 & --     \\
 Oct 18, 2011 & 55853.470 & 10.218 &  9.793 &  8.264 & --     \\
 Oct 31, 2011 & 55866.396 & 10.338 &  9.843 &  8.285 & --     \\
 Nov  3, 2011 & 55869.303 & 10.334 &  9.847 &  8.246 & --     \\
 Nov  8, 2011 & 55874.360 & 10.097 &  9.662 &  8.105 & --     \\
 Nov 13, 2011 & 55879.357 & 10.143 &  9.558 &  7.959 & --     \\
 Nov 15, 2011 & 55881.364 & 10.009 &  9.472 &  7.872 & --     \\
 Nov 16, 2011 & 55882.321 & 10.143 &  9.492 &  7.873 & --     \\
 Nov 17, 2011 & 55883.332 & 10.125 &  9.464 &  7.840 & --     \\
 Nov 18, 2011 & 55884.361 &  9.964 &  9.381 &  7.768 & --     \\
\hline
\hline
\end{tabular}
\end{center}
\normalsize
\end{table*}
\clearpage
%
%
%
\begin{table*}
\begin{center}
\caption{CCD~~$U,B,V,R_{\rm C}$ observations of CH~Cyg 
         from the Star\'a Lesn\'a observatory.}
\begin{tabular}{lccccc}
\hline
\hline
   Date      & JD~24...   & $U$   &  $B$    &  $V$  &$R_{\rm C}$ \\
\hline
 Aug  8, 2007 & 54321.380 & 11.805 & 11.220 &  9.239 &  6.749 \\
 Sep 21, 2007 & 54365.400 & 11.899 & 11.476 &  9.561 & --     \\
 Sep 23, 2007 & 54367.339 & 11.923 & 11.505 &  9.565 & --     \\
 May 10, 2008 & 54597.482 & 11.664 & 10.248 &  8.204 & --     \\
 May 15, 2008 & 54601.592 & 11.618 & 10.159 &  8.107 & --     \\
 May 17, 2008 & 54603.596 & 11.625 & 10.123 &  8.076 & --     \\
 Jun 30, 2008 & 54647.575 & 11.816 & 10.800 &  8.845 & --     \\
\hline
\end{tabular}
\end{center}
\end{table*}
\addtocounter{table}{-1}
\begin{table*}
\begin{center}
\caption{Continued}
\begin{tabular}{lccccc}
\hline
  Date     & JD~24...   &  $U$    &  $B$   & $V$  & $R_{\rm C}$ \\
\hline
 Sep 22, 2009 & 55097.255 & 10.711 &  9.832 &  7.961 & --     \\
 Sep 25, 2009 & 55100.368 & 10.859 &  9.875 &  7.987 & --     \\
 Oct 21, 2009 & 55126.297 &  9.702 &  9.633 &  8.083 &  5.797 \\
 Oct 31, 2009 & 55136.277 &  9.639 &  9.628 &  8.107 & --     \\
 Nov  2, 2009 & 55138.323 &  9.568 &  9.593 &  8.094 &  5.792 \\
 Nov  5, 2009 & 55141.249 &  9.820 &  9.679 &  8.102 &  5.806 \\
 Nov 13, 2009 & 55149.288 &  9.638 &  9.665 &  8.090 &  5.786 \\
 Nov 15, 2009 & 55151.233 &  9.392 &  9.402 &  8.001 &  5.747 \\
 Nov 16, 2009 & 55152.269 &  9.783 &  9.640 &  8.130 &  5.771 \\
 Nov 23, 2009 & 55159.332 &  9.548 &  9.509 &  8.059 &  5.763 \\
 Nov 26, 2009 & 55162.214 & 10.003 &  9.714 &  8.099 & --     \\
 Nov 28, 2009 & 55164.218 & 10.125 &  9.730 &  8.110 & --     \\
 Dec  2, 2009 & 55168.279 & 10.289 &  9.850 &  8.154 &  5.797 \\
 Dec  4, 2009 & 55170.161 & 10.482 &  9.872 &  8.115 &  5.840 \\
 Jan 16, 2010 & 55213.191 & 11.106 & 10.024 &  8.040 &  5.829 \\
 Jan 22, 2010 & 55219.179 & 11.145 & 10.047 &  8.127 &  5.811 \\
 Jan 23, 2010 & 55220.179 & 11.260 &  9.995 &  8.030 &  5.799 \\
 Jan 25, 2010 & 55221.730 & 11.300 &  9.871 &  8.031 & --     \\
 Feb 26, 2010 & 55253.694 & 11.527 & 10.573 &  8.640 &  6.333 \\
 Mar  2, 2010 & 55257.687 & 11.520 & 10.567 &  8.690 &  6.410 \\
 Mar  4, 2010 & 55259.686 & 11.719 & 10.581 &  8.693 &  6.374 \\
 Mar  9, 2010 & 55264.684 & 11.433 & 10.505 &  8.709 &  6.466 \\
 Mar 25, 2010 & 55280.654 & 11.843 & 10.729 &  8.749 & --     \\
 Apr  2, 2010 & 55288.642 & 11.892 & 10.676 &    --  &  6.404 \\
 Apr 20, 2010 & 55306.613 & 11.301 & 10.712 &  8.761 &  6.460 \\
 Apr 25, 2010 & 55311.603 & 11.479 & 10.721 &  8.856 &  6.535 \\
 Apr 29, 2010 & 55315.604 & 11.349 & 10.694 &  8.900 &  6.591 \\
\hline
\hline
\end{tabular}
\end{center}
\normalsize
\end{table*}
\clearpage
%
%
%
\begin{table*}
\begin{center}
\caption{Photoelectric~~$U,B,V,R_{\rm C}$ observations of CI~Cyg 
         from the Skalnat\'e Pleso and Hvar observatories.}
\begin{tabular}{lccccc}
\hline
\hline
 Date & JD~24... & $U$ &  $B$   &  $V$ &$\Delta R_{\rm C}$ \\   
\hline
 Feb 15, 2007 & 54146.569 & 11.331 & 11.833 & 10.731 &  0.515 \\
 Mar 13, 2007 & 54172.557 & 11.408 & 11.941 & 10.717 &  0.515 \\
 Mar 14, 2007 & 54173.561 & 11.429 & 11.914 & 10.719 &  0.555 \\
 Apr 13, 2007 & 54203.541 & 11.384 & 11.896 & 10.656 &  0.510 \\
 Jun  4, 2007 & 54256.484 & 11.351 & 11.860 & 10.708 &  0.533 \\
 Jun 16, 2007 & 54268.490 & 11.477 & 11.955 & 10.754 &  0.591 \\
 Jul  2, 2007 & 54283.525 & 11.476 & 11.942 & 10.763 &  0.593 \\
\hline
\end{tabular}
\end{center}
\end{table*}
\addtocounter{table}{-1}
\begin{table*}
\begin{center}
\caption{Continued}
\begin{tabular}{lccccc}
\hline
  Date   & JD~24...   &  $U$   & $B$   & $V$  &$\Delta R_{\rm C}$ \\
\hline
 Apr 29, 2010 & 55316.483 & 11.711 & 11.993 & 10.797 & --     \\
 Apr 30, 2010 & 55317.423 & 11.684 & 12.010 & 10.743 & --     \\
 Jun  9, 2010 & 55357.379 & 11.682 & 12.098 & 10.857 & --     \\
 Jun 29, 2010 & 55377.488 & 11.666 & 11.915 & 10.636 & --     \\
 Jul  9, 2010 & 55387.448 & 11.380 & 11.778 & 10.607 & --     \\
 Jul 10, 2010 & 55387.500 &  --    & 11.705 & 10.528 & --     \\
 Jul 22, 2010 & 55400.393 & 11.252 & 11.699 & 10.612 & --     \\
\hline
\end{tabular}
\end{center}
\end{table*}
\addtocounter{table}{-1}
\begin{table*}
\begin{center}
\caption{Continued}
\begin{tabular}{lccccc}
\hline
  Date   & JD~24...   &  $U$   & $B$   & $V$  &$\Delta R_{\rm C}$ \\
\hline
 Oct  3, 2011 & 55838.400 & 11.242 & 11.766 & 10.690 & --     \\
 Oct  9, 2011 & 55844.320 &   --   & 11.848 & 10.770 & --     \\
 Oct 17, 2011 & 55852.421 & 11.267 & 11.790 & 10.750 & --     \\
 Oct 31, 2011 & 55866.349 & 11.322 & 11.852 & 10.693 & --     \\
 Nov  3, 2011 & 55869.261 & 11.277 & 11.838 & 10.657 & --     \\
 Nov  6, 2011 & 55872.262 & 11.394 & 11.858 & 10.714 & --     \\
 Nov  8, 2011 & 55874.234 & 11.299 & 11.870 & 10.645 & --     \\
 Nov 13, 2011 & 55879.288 & 11.285 & 11.852 & 10.693 & --     \\
 Nov 15, 2011 & 55881.321 & 11.326 & 11.793 & 10.679 & --     \\
 Nov 16, 2011 & 55882.268 & 11.229 & 11.853 & 10.678 & --     \\
 Nov 17, 2011 & 55883.281 & 11.355 & 11.875 & 10.702 & --     \\
 Nov 18, 2011 & 55884.254 & 11.320 & 11.884 & 10.657 & --     \\
\hline
\hline
\end{tabular}
\end{center}
$\dagger$ -- Observations obtained at the Hvar observatory 
\normalsize
\end{table*}
%
%
%
\begin{table*}
\begin{center}
\caption{CCD~~$U,B,V,R_{\rm C},I_{\rm C}$ observations of CI~Cyg 
         from the Star\'a Lesn\'a observatory.}
\begin{tabular}{lcccccc}
\hline
\hline
  Date   & JD~24...   & $U$  &  $B$  &  $V$  &$R_{\rm C}$&$I_{\rm C}$  \\
\hline
 Jul 23, 2007 & 54305.381 & 11.428 & 11.935 & 10.732 &  9.303 &  7.708 \\
 Sep 20, 2007 & 54364.386 & 11.495 & 12.029 & 10.794 &  9.365 &  7.657 \\
 Sep 21, 2007 & 54365.439 & 11.459 & 12.002 & 10.727 &  9.314 &  7.675 \\
 Sep 23, 2007 & 54367.373 & 11.508 & 12.053 & 10.801 &  9.361 &  7.662 \\
 Oct 10, 2007 & 54384.279 & 11.518 & 12.028 & 10.811 &  9.373 &  7.666 \\
 Dec 20, 2007 & 54455.216 & 11.874 & 12.308 & 11.201 &  9.723 &  7.886 \\
 May 12, 2008 & 54598.557 & 11.646 & 12.126 & 10.875 &  9.435 &  7.627 \\
\hline
\hline
\end{tabular}
\end{center} 
\normalsize  
\end{table*} 
\clearpage
%
%
\begin{table*}
\begin{center}
\caption{CCD~~$I_{\rm C},R_{\rm C},V,B,U$ observations
         from the Rozhen Observatory.}
\begin{tabular}{lccccccl}
\hline
\hline
 Date   & JD~24... &$I_{\rm C}$& $R_{\rm C}$& $V$ & $B$ & $U$ & Telescope \\
\hline \\[-2.5mm]
\multicolumn{8}{c}{EG~And} \\
\hline
Nov 17, 2006&   54057.356& --   & --   &  7.283&  8.898&  --    & Schm\\
Dec 16, 2006&   54086.356& --   & --   &  7.246&  8.886&  --    & Schm\\
Feb 29, 2008&   54256.340& --   & --   &  7.166&  8.745&  10.407& Schm\\
Nov 20, 2008&   54791.496& 5.553& 6.694&  7.524&  8.834&  --    & Schm\\
Jan 11, 2009&   54843.292& 5.364& 6.496&  7.222&  8.732&  --    & Schm\\
\hline
\end{tabular}
\end{center}
\end{table*}
\addtocounter{table}{-1}
\begin{table*}
\begin{center}
\caption{Continued}
\begin{tabular}{lccccccl}
\hline
 Date   & JD~24... &$I_{\rm C}$& $R_{\rm C}$& $V$ & $B$ & $U$ & Telescope \\
\hline
Jun 24, 2009&	55007.581&10.466&11.580&13.060&	13.642&--     &	1.3m RC	\\	
Jul 06, 2009&	55019.500&10.421&11.666&13.100&	13.757&--     &	1.3m RC	\\
Jul 14, 2009&	55027.381&10.498&11.743&13.151&	13.795&--     &	Schm	\\	
Jul 16, 2009&	55029.391&10.512&11.768&13.177&	13.807&--     &	Schm	\\	
Jul 27, 2009&	55040.418&10.528&11.789&13.229&	13.865&	13.263&	1.3m RC	\\	
Aug 21, 2009&	55065.266&10.538&11.905&13.356&	14.025&--     &	Schm	\\	
Oct 08, 2009&	55113.229&10.580&12.266&13.621&	14.319&--     &	Schm	\\
\hline
\end{tabular}
\end{center}
\end{table*}
\addtocounter{table}{-1}
\begin{table*}
\begin{center}
\caption{Continued}
\begin{tabular}{lccccccl}
\hline
 Date   & JD~24... &$I_{\rm C}$& $R_{\rm C}$& $V$ & $B$ & $U$ & Telescope \\
\hline \\[-2.5mm]
\multicolumn{8}{c}{AX~Per} \\
\hline 
Nov 17, 2006& 54057.367& --   & --    &  11.082&  12.385&  --    &  Schm\\
Dec 16, 2006& 54086.368& --   & --    &  11.085&  12.372&  --    &  Schm\\
Feb 29, 2008& 54526.349& --   & --    &  11.620&  12.757&  12.647&  Schm\\
Nov 20, 2008& 54791.504& 8.687& 10.338&  11.618&  12.499&  --    &  Schm\\
Jan 11, 2009& 54843.300& 8.788& 10.542&  11.758&  12.630&  --    &  Schm\\
Jul 15, 2009& 55028.555& 8.355& 10.400&  11.603&  12.792&  --    &  Schm\\
Oct 08, 2009& 55113.459& 8.621& 10.352&  11.305&  11.983&  --    &  Schm\\
Nov 21, 2009& 55157.419& 8.517& 10.250&  11.318&  12.048&  --    &  Schm\\
\hline
\hline 
\end{tabular}
\end{center}
1)~Schm -- 50/70 cm Schmidt,
\normalsize 
\end{table*}
%
%
%
\begin{table*}
\begin{center}
\caption{Photoelectric~~$U,B,V,R_{\rm C}$ observations of TX~CVn 
         from the Skalnat\'e Pleso observatory.}
\begin{tabular}{lccccc}
\hline
\hline
 Date & JD~24... & $U$ &  $B$   &  $V$ &$\Delta R_{\rm C}$ \\   
\hline
 Feb 14, 2007 & 54146.417 & 10.410 & 10.342 &  9.782 &  0.037 \\
 Mar 11, 2007 & 54171.378 & 10.564 & 10.481 &  9.870 &  0.109 \\
 Mar 12, 2007 & 54172.365 & 10.543 & 10.506 &  9.867 &  0.127 \\
 Mar 13, 2007 & 54173.353 & 10.495 & 10.500 &  9.876 &  0.126 \\
 Mar 25, 2007 & 54185.350 &  --    & 10.501 &  9.855 &  0.088 \\
 Apr  4, 2007 & 54195.320 & 10.517 & 10.500 &  9.895 &  0.162 \\
 Apr 11, 2007 & 54202.296 & 10.513 & 10.527 &  9.885 &  0.141 \\
 Apr 14, 2007 & 54205.358 & 10.508 & 10.505 &  9.884 &  0.139 \\
 Nov  3, 2007 & 54407.555 & 10.543 & 10.508 &  9.876 &  0.104 \\
 Nov  6, 2007 & 54410.533 & 10.513 & 10.492 &  9.909 &  0.116 \\
 Dec 17, 2007 & 54451.658 & 10.539 & 10.534 &  9.887 &  0.123 \\
 Dec 26, 2007 & 54460.694 & 10.546 & 10.517 &  9.871 &  0.137 \\
 Dec 26, 2007 & 54461.461 & 10.600 & 10.561 &  9.906 &  0.123 \\
 Dec 28, 2007 & 54462.507 & 10.528 & 10.437 &  9.842 &  0.067 \\
 Feb 24, 2008 & 54521.371 & 10.656 & 10.568 &  9.954 & --     \\
 Feb 25, 2008 & 54522.434 & 10.721 & 10.560 &  9.953 & --     \\
 May  7, 2008 & 54594.434 & 10.771 & 10.682 &  9.997 & --     \\
\hline
\hline 
\end{tabular}
\end{center}
\normalsize 
\end{table*}
\clearpage
%
%
%
\begin{table*}
\begin{center}
\caption{Photoelectric~~$U,B,V,R_{\rm C}$ observations of AG~Dra 
         from the Skalnat\'e Pleso observatory.}
\begin{tabular}{lccccc}
\hline
\hline
 Date & JD~24... & $U$ &  $B$   &  $V$ &$\Delta R_{\rm C}$ \\
\hline
 Mar 11, 2007 & 54171.332 &  9.162 &  9.998 &  9.247 & -1.055 \\
 Mar 12, 2007 & 54172.316 &  9.152 &  9.985 &  9.217 & -1.115 \\
 Mar 13, 2007 & 54173.308 &  9.129 &  9.933 &  9.191 & -1.075 \\
 Mar 25, 2007 & 54185.320 &   --   & 10.136 &  9.186 & -1.010 \\
 Apr  4, 2007 & 54195.391 &  9.187 & 10.025 &  9.244 & -1.055 \\
 Apr 11, 2007 & 54202.329 &  9.243 & 10.053 &  9.241 & -1.052 \\
 Apr 12, 2007 & 54203.323 &  9.260 & 10.060 &  9.257 & -1.041 \\
\hline
\end{tabular}
\end{center}
\end{table*}
\addtocounter{table}{-1}
\begin{table*}
\begin{center}
\caption{Continued}
\begin{tabular}{lccccc}
\hline
  Date   & JD~24...   &  $U$   & $B$   & $V$  &$\Delta R_{\rm C}$ \\
\hline
 Nov 22, 2009 & 55157.640 & 11.625 & 11.164 &  9.836 & --     \\
 Jan 16, 2010 & 55212.576 & 11.462 & 11.134 &  9.786 & --     \\
 Feb 27, 2010 & 55255.448 & 11.610 & 11.240 &  9.870 & --     \\
 Mar  9, 2010 & 55265.467 & 11.579 & 11.110 &  9.765 & --     \\
 Mar 18, 2010 & 55274.390 & 11.545 & 11.107 &  9.763 & --     \\
 Mar 22, 2010 & 55278.406 & 11.604 & 11.196 &  9.800 & --     \\
 Mar 24, 2010 & 55280.323 & 11.503 & 11.159 &  9.828 & --     \\
\hline
\hline
\end{tabular}
\end{center}
\normalsize 
\end{table*}
\clearpage
%
%
%
\begin{table*}
\begin{center}
\caption{CCD~~$U,B,V,R_{\rm C},I_{\rm C}$ observations of AG~Dra 
         from the Star\'a Lesn\'a observatory.}
\begin{tabular}{lcccccc}
\hline
\hline
  Date   & JD~24...   & $U$  &  $B$  &  $V$  &$R_{\rm C}$&$I_{\rm C}$  \\
\hline
 Mar  3, 2005 & 53432.504 & 11.381 & 11.338 &  9.883 &  9.181 &  8.366 \\
 May 21, 2005 & 53512.335 & 11.300 & 11.287 &  9.788 &  8.929 &  8.258 \\
 May 25, 2005 & 53516.338 & 11.276 & 11.306 &  9.782 &  8.897 &  8.250 \\
 May 26, 2005 & 53517.337 & 11.292 & 11.306 &  9.769 &  8.883 &  8.237 \\
 Jul 20, 2005 & 53572.422 &  9.423 &  --    &   --   &  8.524 &  7.870 \\
 Jul 24, 2005 & 53576.335 &   --   & 10.375 &  9.311 &  8.568 &  7.943 \\
 Jul 27, 2005 & 53579.367 &  9.526 & 10.333 &  9.276 &  8.494 &  7.913 \\
\hline
\hline
\end{tabular}
\end{center} 
\normalsize  
\end{table*} 
\clearpage
%
%
%
\begin{table*}
\begin{center}
\caption{Photoelectric~~$U,B,V,R_{\rm C}$ observations of AG~Peg 
         from the Skalnat\'e Pleso observatory.}
\begin{tabular}{lccccc}
\hline
\hline
 Date & JD~24... & $U$ &  $B$   &  $V$ &$\Delta R_{\rm C}$ \\   
\hline
 Jul 15, 2007 & 54297.419 &  9.834 &  9.899 &  8.639 & -0.020 \\
 Jul 17, 2007 & 54299.457 &  9.694 &  9.906 &  8.638 & --     \\
 Aug 26, 2007 & 54338.546 &  9.890 &  9.718 &  8.517 & -0.059 \\
 Sep 14, 2007 & 54357.511 & 10.024 & 10.018 &  8.729 &  0.072 \\
 Oct 14, 2007 & 54388.464 & 10.131 & 10.052 &  8.728 &  0.095 \\
 Nov  5, 2007 & 54410.318 & 10.204 & 10.151 &  8.896 &  0.250 \\
 Dec 18, 2007 & 54453.214 & 10.265 & 10.154 &  8.818 &  0.147 \\
\hline
\hline
\end{tabular}
\end{center} 
\normalsize  
\end{table*} 
\clearpage
%
%
\begin{table*}
\begin{center}
\caption{Photoelectric~~$U,B,V,R_{\rm C}$ observations of AX~Per 
         from the Skalnat\'e Pleso observatory.}
\begin{tabular}{lccccc}
\hline
\hline
 Date & JD~24... & $U$ &  $B$   &  $V$ &$\Delta R_{\rm C}$ \\   
\hline
 Mar 12, 2007 & 54172.266 & 12.420 & 13.012 & 11.954 &  3.705 \\
 Mar 25, 2007 & 54185.272 & 12.350 & 12.964 & 11.780 &  3.440 \\
 Jul 15, 2007 & 54297.463 & 11.674 & 12.122 & 11.161 &  3.041 \\
 Jul 17, 2007 & 54299.489 & 11.469 & 12.076 & 11.094 & --     \\
 Jul 26, 2007 & 54308.498 & 11.469 & 12.033 & 11.061 & --     \\
 Aug  3, 2007 & 54315.529 & 11.477 & 12.068 & 11.098 &  3.000 \\
 Aug 26, 2007 & 54338.600 & 12.602 & 12.890 & 11.540 &  3.460 \\
\hline
\end{tabular}
\end{center}
\end{table*}
\addtocounter{table}{-1}
\begin{table*}
\begin{center}
\caption{Continued}
\begin{tabular}{lccccc}
\hline
  Date   & JD~24...   &  $U$   & $B$   & $V$  &$\Delta R_{\rm C}$ \\
\hline
 Feb 21, 2011 & 55614.291 & 11.372 & 12.015 & 11.035 & --     \\
 Feb 27, 2011 & 55620.326 & 11.296 & 11.982 & 11.088 & --     \\
 Mar  7, 2011 & 55628.328 & 11.451 & 12.000 & 11.073 & --     \\
 Sep 26, 2011 & 55831.496 & 12.498 & 12.851 & 11.616 & --     \\
 Oct  1, 2011 & 55835.627 & 12.422 & 12.759 & 11.496 & --     \\
 Oct 19, 2011 & 55853.642 &   --   & 12.730 & 11.483 & --     \\
 Nov  1, 2011 & 55866.639 & 12.397 & 12.767 & 11.547 & --     \\
 Nov  6, 2011 & 55872.416 & 12.321 & 12.729 & 11.524 & --     \\
 Nov  8, 2011 & 55873.544 & 12.434 & 12.741 & 11.516 & --     \\
 Nov 16, 2011 & 55881.551 & 12.267 & 12.699 & 11.459 & --     \\
 Nov 17, 2011 & 55882.555 & 12.289 & 12.673 & 11.461 & --     \\
 Nov 18, 2011 & 55883.566 & 12.315 & 12.644 & 11.432 & --     \\
 Nov 19, 2011 & 55884.617 & 12.252 & 12.660 & 11.417 & --     \\
\hline
\hline
\end{tabular}
\end{center} 
\normalsize  
\end{table*} 
\end{document}